\RequirePackage{fix-cm}
\pdfoutput=1
\documentclass[natbib,smallextended]{svjour3}       
\smartqed  
\usepackage{graphicx}
\usepackage{amsmath}
\usepackage{amssymb}
\usepackage{aas_macros}
\newcommand{\msunh}{h^{-1}M_{\odot}}
\newcommand{\msun}{M_{\odot}}
\newcommand{\be}{\begin{equation}}
\newcommand{\ee}{\end{equation}}
\newcommand{\ASCA}{{\it ASCA}}
\newcommand{\Beppo}{{\it Beppo-SAX}}
\newcommand{\XMM}{{\it XMM-Newton}}
\newcommand{\Chandra}{{\it Chandra}}
\newcommand{\Suzaku}{{\it Suzaku}}
\newcommand{\XRISM}{{\it XRISM}}
\newcommand{\Athena}{{\it Athena}}
\newcommand{\Hitomi}{{\it Hitomi}}
\newcommand{\rvir}{R_{vir}}
\newcommand{\rfive}{R_{500}}
\newcommand{\rtwo}{R_{200}}
\newcommand{\Mrev}{Mernier et al., subm., this issue}
\journalname{SSRv}

\begin{document}

\title{Enrichment of the hot intracluster medium: numerical simulations}

\author{V. Biffi \and F. Mernier \and P. Medvedev}

\institute{V. Biffi \at
  Physics Department, Astronomy Unit, Trieste University, v. Tiepolo 11, 34143 Trieste (Italy) \\
  INAF, Observatory of Trieste, v. Tiepolo 11, 34143 Trieste (Italy)\\
  \email{biffi@oats.inaf.it}\\
  \and
  F. Mernier \at
  MTA-E\"otv\"os University Lend\"ulet Hot Universe Research Group,
  P\'azm\'any P\'eter s\'et\'any 1/A, Budapest, 1117, Hungary \\
  Institute of Physics, E\"otv\"os University, P\'azm\'any P\'eter s\'et\'any 1/A, Budapest, 1117, Hungary \\
  SRON Netherlands Institute for Space Research, Sorbonnelaan 2, 3584 CA Utrecht, The Netherlands \\
  \and
  P. Medvedev \at
  Space Research Institute of the Russian Academy of Sciences (IKI), 84/32
Profsoyuznaya Str, Moscow, 117997, Russia
}

\date{Received: date / Accepted: date}

\maketitle

\begin{abstract}
  The distribution of chemical elements in the hot intracluster medium
  (ICM) retains valuable information about the enrichment and star
  formation histories of galaxy clusters, and on the feedback and
  dynamical processes driving the evolution of the cosmic baryons.
  In the present study we review the progresses made so far in the
  modelling of the ICM chemical enrichment in a cosmological context,
  focusing in particular on cosmological hydrodynamical simulations.
  We will review the key aspects of embedding chemical evolution
  models into hydrodynamical simulations, with special attention to
  the crucial assumptions on the initial stellar mass function,
  stellar lifetimes and metal yields, and to the numerical limitations
  of the modelling.
  At a second stage, we will overview the main simulation results
  obtained in the last decades and compare them to X-ray observations of
  the ICM enrichment patterns. In particular, we will discuss how
  state-of-the-art simulations are able to reproduce the observed
  radial distribution of metals in the ICM, from the core to the
  outskirts, the chemical diversity depending on cluster
  thermo-dynamical properties, the evolution of ICM metallicity and
  its dependency on the system mass from group to cluster scales.
  Finally, we will discuss the limitations still present in modern
  cosmological, chemical, hydrodynamical simulations and the
  perspectives for improving the theoretical modelling of the ICM
  enrichment in galaxy clusters in the future.
  \keywords{Galaxy Clusters \and ICM chemical enrichment \and Numerical Simulations}
\end{abstract}

\section{Introduction}
\label{intro}

In the hierarchical paradigm of structure formation, clusters of
galaxies are the largest and latest systems formed in the Universe
from the gravitational collapse of large overdense regions.  Despite a
dominating dark matter component ($\sim 85\%$ of the total mass
budget), observations of galaxy clusters reveal that a fraction of
their content is in the form of stars in galaxies and of a hot diffuse
gas, the intra-cluster medium (ICM).
Filling the dark-matter dominated potential wells of galaxy clusters
and groups, this hot plasma carries the imprint of the cluster
formation and evolution, and the study of its thermodynamical and
chemical properties is an invaluable source of information about the
star formation and feedback histories.

X-ray observations of the hot ($T\sim 10^7\mbox{--}10^8$\,K)
ICM allow us to derive directly its thermal properties (from density
and temperature), but have also shown that the ICM spectrum is
characterized by emission lines from heavy elements. This indicates that a
non-negligible fraction of this gas must have been enriched with heavy
ions from stars (principally) within galaxies before it became part of
the diffuse ICM component.
The abundance and spatial distribution of these metals (namely all the
chemical elements heavier than helium) is therefore influenced by
several complex physical and dynamical processes that drive the formation
and evolution of galaxies and their interaction with
the ambient intracluster medium~\cite[see][for a review]{schindler2008}.
From the study of the gas chemical enrichment in local clusters
and its evolution with time, we can therefore derive
constraints on the main physical processes shaping their observable
properties, from the merging and star-formation histories,
to the effects of non-gravitational sources of energy
like feedback from supernova (SN) winds and Active Galactic Nuclei (AGN).

The metals that enrich the ICM are produced by stars and stellar remnants.
In particular, $\alpha$- and iron-peak elements are mostly
synthetized during stellar lifetimes and SN explosions, and are then
released into the surrounding gas.
Core-collapse supernovae (to which we refer as SNII, in the
following),
originate from massive short-living stars and
primarily produce oxygen (O) and some other
$\alpha$-elements, such as neon (Ne) and magnesium (Mg).
Type-Ia supernovae (SNIa), arising from the explosions of white dwarf
stars in binary systems, are mostly responsible for the production of
heavier elements, especially iron (Fe) and nickel (Ni).
Intermediate-mass elements like silicon (Si), sulfur (S), argon (Ar)
and calcium (Ca) are mainly released by SNII, whereas the
  bulk of carbon (C) and nitrogen (N) and neutron-rich isotopes of Ne,
  Mg, as well as Na, Al are produced and ejected by low- and
  intermediate-mass stars during the asymptotic giant branch (AGB)
  phase~\cite[e.g.][]{boehringer2010,nomoto2013,deplaa2013,karakas2014}.
Despite this picture is fairly well established, several uncertainties
still remain, especially in the details of stellar evolution models
for SN. Thus, useful constraints on the various models can be derived
from measurements of abundance patterns of different elements in the
intracluster gas, and from differences with respect to the theoretical
model predictions (see \Mrev, for a recent detailed review
on chemical measurements from X-ray data).

Other than observations of the metal abundances in the ICM,
theoretical investigations by means of numerical simulations are
extremely powerful
to explore the enrichment history of the gas.  In fact,
  cosmological hydrodynamical simulations uniquely allow to
  simultaneously follow a large variety of highly non-linear, complex
  physical processes driving the co-evolution of dark matter and
  baryonic matter in the Universe.  In particular, state-of-the-art
  simulations can account for detailed treatments of the most
  important gas-dynamical processes that shape the properties of
  galaxy clusters during their formation and evolution: from gas
  cooling and star formation, to stellar evolution and consequent
  production of chemical elements, including stellar and AGN feedback
  processes.
 Therefore, they can be exploited to investigate the details of the
  ICM chemical properties at different epochs, the expected
signatures of energy feedback processes on the distribution of metals,
and the relation between the distribution of chemical elements in the
ICM and the cluster formation and evolution.
Moreover, simulations can be used to study the connection between
different enrichment patterns of the ICM and the resulting
thermal structure and star formation history of galaxy clusters,
given that metals alter the gas cooling through their several cooling~lines.

Since the late 90s, several groups started to couple chemical
evolution models with hydrodynamical simulations of
galaxies~\cite[][]{raiteri1996,mosconi2001}, and of galaxy
clusters~\cite[][]{lia2002,kawata2003,valdarnini2003,tornatore2004,scannapieco2005,tornatore2007,dave2008,wiersma_chem2009}.
In the past 15 years, advancements in the modelling of the chemical
properties have also been accompanied by substantial improvements in
the description of the various physical processes driving the
evolution of the baryonic component, such has energy feedback from
stellar and AGN
sources~\cite[][]{springeldimatteo2005,dimatteo2005,scannapieco2005,sijacki2006,fabjan2010,teyssier2011,dubois2011,planelles2014,martizzi2016,barnes2017}.
Despite some limitations still persist (e.g.\ on the details of the baryonic
cycle in cluster cores and of the feedback scheme), cluster
simulations have reached an unprecedented level of detail.
Recent simulation campaigns carried on by different groups have in
fact succeeded in reproducing more realistic cluster properties, for instance recovering the observed thermal and chemical diversity of cool-core and non-cool-core clusters~\cite[][]{rasia2015,hahn2015,martizzi2016,biffi2017,barnes2018,vogelsberger2018}.
Although not discussed in this paper, theoretical investigations of
the chemical enrichment of the inter-galactic gas in clusters and
groups have also been pursued via semi-analytical techniques, coupling
DM-only or non-radiative simulations with semi-analytical models of
galaxy formation and evolution that account, with various level of
detail, for the chemical enrichment processes from SN explosions and
AGB stars in
galaxies. Several of these works, explored the spatial distribution of
metals in the ICM and the impact of the various physical and dynamical processes onto the final cluster enrichment level~\cite[][]{delucia2004,nagashima2005,schindler2005,domainko2006,cora2006,cora2008,moll2007,kapferer2007,short2013,yates2017}.

With respect to the review by~\cite{borgani2008} on ICM chemical
  enrichment in simulated clusters, important improvements have been
  obtained from the numerical side. Most of all, the difficulties of
  regulating overcooling at the cluster center while reproducing the
  thermodynamical diversity of observed cool-core and non-cool-core
  systems have been overcome in the past five years, as discussed
  above. This allowed different independent groups to obtain more and
  more realistic simulations of galaxy clusters, in which to explore
  more faithfully the details of the ICM chemo-dynamical history.
  Therefore, we review in this article the latest and most important
results on the chemical enrichment of the ICM, focusing on
  state-of-the-art numerical hydrodynamical simulations in a
cosmological context.

In particular, the paper is structured as follows. In
Sec.~\ref{sec:1} we outline the building blocks of chemical evolution
models and overview the progresses made so far to include them into
cosmological hydrodynamical simulations (Sec.~\ref{sec:chem-sims}). In
Sec.~\ref{sec:3} we summarize the most important simulation results on
the modelling of the distribution of chemical elements in the ICM,
from the centermost regions out to the outskirts, while its evolution
with time is discussed in Sec.~\ref{sec:evol}.  The dependence of the
chemical enrichment on the system scale, from groups to clusters of
galaxies, is then reviewed in Sec.~\ref{sec:clus-grp}.
Finally, we conclude by overviewing the principal limitations that are
still present in modern cosmological chemo- and hydro-dynamical
simulations (Sec.~\ref{sec:concl}).

As such, this article complements the review by Mernier et
al. 2018 (subm., this issue), where the most significant achievements
in measuring the ICM chemical properties from X-ray observations are
reviewed.

\section{Chemical evolution models in simulations}\label{sec:1}

In this section, we provide a brief description of the basic structure
and features of chemical evolution models implemented in
numerical simulations~\cite[e.g., see also][]{borgani2008}.
Including these models in simulations of galaxies and clusters is of
great importance because the metal-rich ejecta from evolved stellar
populations determine the heavy-element content of the
inter-galactic and intra-cluster gas.

Cosmological hydrodynamical simulations of large volumes, even in the
case of {\it zoom-in} re-simulations of galaxies or galaxy clusters,
cannot reach the resolution needed to resolve single stars and must
therefore resort to sub-grid models of star formation and chemical
enrichment. Typically, star formation is described by the conversion
of a gas element into a ``star'' element, as a consequence of the
cooling of the gas.
This approach was adopted, for instance, in the first implementation of
star formation in SPH simulations by~\cite{katz1992},
and later in the effective models for multi-phase inter-stellar medium description
presented by~\cite{springel2003} or
by~\cite{MUPPI2010}. Given the typical resolution of large-scale
simulations, the star element has a mass of $10^4\mbox{--}10^6\msunh$
and rather represents a population of stars, all with the same age and
initial metallicity, i.e.\ a simple stellar population (SSP),
characterized by an assumed initial mass function (IMF).
For every star element in the simulation one can compute the
evolution of the stellar population that it represents, by assuming
not only the IMF, but also some mass-dependent lifetime function
and stellar yields associated to the various mass ranges and enrichment channels.

Briefly, we recall hereafter these pillars of chemical evolution
models embedded into cosmological hydrodynamical simulations.

\subsubsection*{Ejection channels of heavy elements}
The chemical feedback in simulations typically
originates from three main stellar evolution channels: SNII, SNIa, and
stars undergoing the AGB
phase~\cite[e.g.][]{tornatore2004,tornatore2007,dave2008,wiersma_chem2009}.

The calculations constituting the core of chemical evolution models are a set of integral equations. These compute the rates at which stars of different masses explode as SNII, SNIa or undergo the AGB phase in each SSP and the corresponding metal ejecta that will enrich the surrounding gas~\cite[we refer the reader to][for a more
  detailed description]{matteucci2003}.

\subsubsection*{Stellar yields}
Stellar yields from stellar evolution models provide the amount of
different metal species released during the evolution of a SSP. This
quantity depends on the star initial mass and metallicity.
Many groups in the past decades have calculated and proposed various
sets of stellar yields for massive, intermediate- and low-mass stars.
In the literature, the most widely used in hydrodynamical
  simulations embedding chemical enrichment models are those
by~\cite{WW1995}, \cite{portinari1998}, \cite{chieffi2004} and
\cite{kobayashi2006} for SNII; the sets proposed by~\cite{nomoto1997},
\cite{Iwamoto1999}, \cite{Thielemann2003} and \cite{travaglio2004} for
SNIa; and for low- and intermediate-mass stars the tables
by~\cite{vandehoek1997}, \cite{marigo2001}, \cite{karakas2007},
\cite{karakas2010} and \cite{doherty2014}~\cite[see also][for
    further references of AGB yields sets available in the
    literature]{karakas2014}.

Unfortunately, yields for many elements (especially the iron-peak
ones) can actually vary significantly, by up to some order of
magnitudes, from author to author. The choice of the set of yields
that is included as an input parameter in the simulations inevitably
affects the final predictions from theoretical models. In fact, this
still represents an important source of uncertainties in predicting
the precise chemical enrichment level of the inter-stellar medium and
ICM derived from cosmological hydrodynamical simulations.

\subsubsection*{Lifetime function}
In order to describe the evolution of a SSP and the enrichment
channels through which the stellar population produces and releases
metals, it is crucial to know the typical lifetimes of stars with
different masses.
Various functional forms for the mass-dependence of
the lifetimes have been proposed in the literature. The most commonly
adopted ones in simulations are those by~\cite{maeder1989},
\cite{padovani1993}, and \cite{chiappini1997}, among which the main
differences concern the lifetimes of low-mass stars.
Typically, the lifetimes are a decreasing function of the stellar mass.
With respect to the enrichment channels considered in simulations, we
note therefore that SNIa and AGB sources are typically related to the
evolution of low- and intermediate-mass stars with longer lifetimes,
whereas viceversa SNII events originate from massive short-living
stars.

In the majority of the cases the above-mentioned mass-dependent
lifetime functions are adopted.
We recall nonetheless that the stellar lifetimes
can also depend on metallicity, like those presented by~\cite{portinari1998} \cite[see also][for an overview and differences between various functional forms in the literature]{romano2005_I}.

\subsubsection*{Initial mass function (IMF)}
Another key ingredient of chemical models is the IMF, namely a functional
form that parametrizes the number of stars as a function of their
mass. Its shape essentially constrains the relative ratio
between low-mass long-living stars and massive short-living ones~\cite[e.g.,][]{romano2005_I}.
The choice of the IMF function has profound impact on the modelling of
metal enrichment since it is directly connected to the relative
abundance of SNIa and SNII, and consequently to the abundance ratios
of elements produced by the two different channels (i.e.\ $\alpha$
vs. Fe-peak elements).
Other than chemical feedback, also the modelling of the stellar energy
feedback, associated to SNII-driven winds, will be affected by the
relative fractions of stars with different masses and specifically to
the abundance of massive stars.

\begin{figure}
  \centering
  \includegraphics[width=0.85\textwidth]{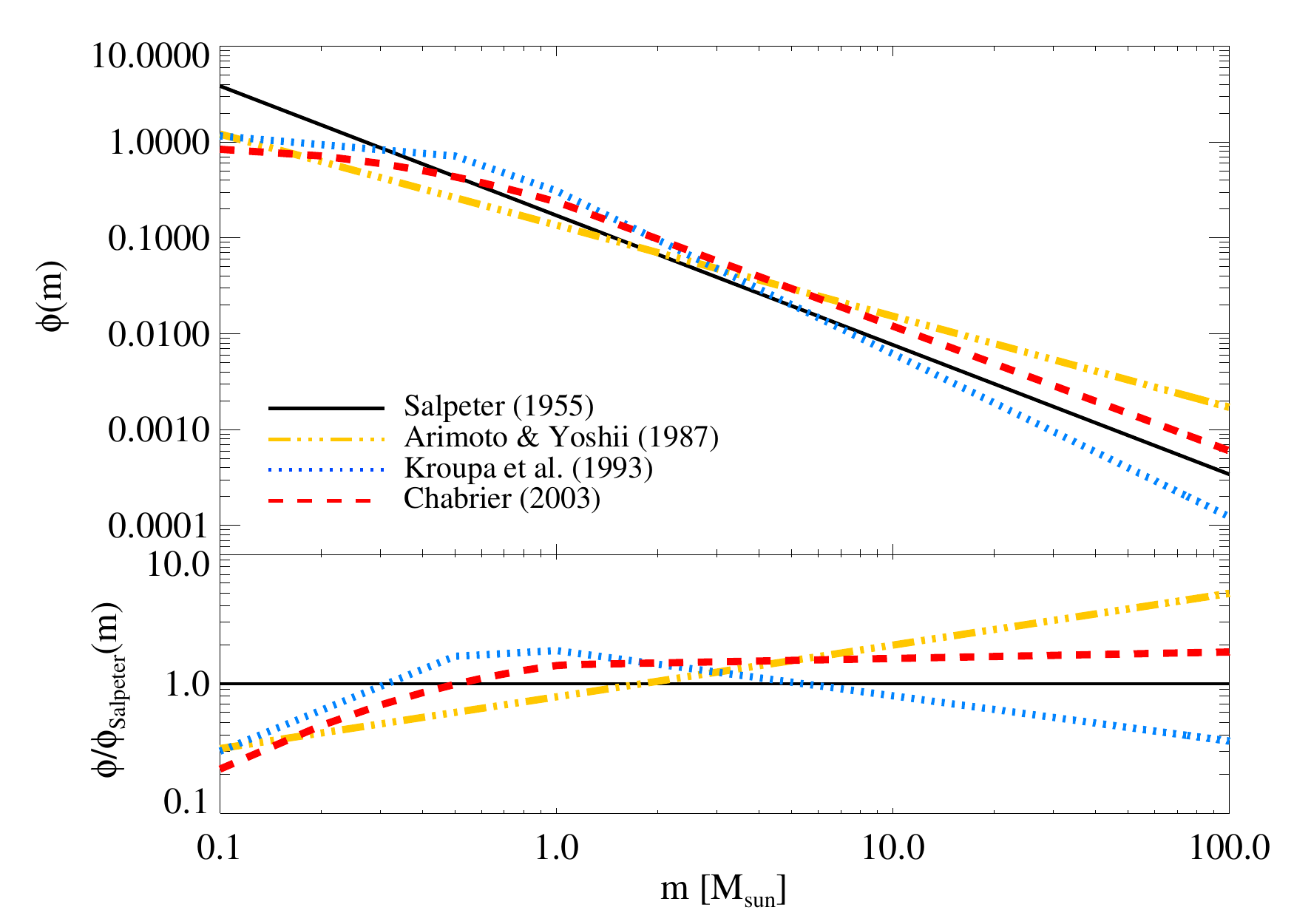}
  \caption{Comparison between different stellar IMFs. As an example we
    show the single power-law IMF functions by~\cite{salpeter1955}
    and~\cite{arimoto1987}, the
    multi-slope power-law IMF by~\cite{kroupa1993}, and the IMF
    proposed by~\cite{chabrier03} with its lognormal shape at
    low stellar masses.
\label{fig:IMFs}}
\end{figure}

\sloppy{Historically, the most commonly adopted IMF is the one
  introduced by~\cite{salpeter1955}, which is characterized by a
  single power-law shape, $\phi(m) \propto m^{-\alpha}$, with a
  (logarithmic) slope of $\alpha=1.35$.}
\cite{arimoto1987} also proposed a single power-law IMF but
with a shallower slope, $\alpha=0.95$, that predicts a larger number of
massive stars compared to the Salpeter IMF. IMF functions of this sort
are typically referred to as top-heavy.

Other IMF functions, also widely used in the literature, allow for a change of the IMF accordingly to the mass. This is the case for the Kroupa
IMF~\cite[][]{kroupa1993,kroupa2001} and the Chabrier
IMF~\cite[][]{chabrier03}. The former is essentially parametrized by a multi-slope power-law form, whereas the Chabrier IMF is characterized by a power law at high stellar masses and by a log-normal
form~\cite[first introduced by][]{millerscalo1979} at the lower-mass
end.  The changes to the single-slope Salpeter IMF were
developed starting from the late 1970s in order to reproduce the
shallower IMF towards the low stellar masses indicated by a number of
observational constraints.

The impact of changing the IMF on the fraction of stars in the various
mass bins can be seen for instance in Fig.~\ref{fig:IMFs}, where we
compare directly the IMFs by~\cite{salpeter1955} ($\phi(m) \propto m^{-1.35}$),
\cite{arimoto1987} ($\phi(m) \propto m^{-0.95}$),
  \cite{kroupa1993} (with 3 slopes: $\alpha=0.3$ for $m< 0.5\,\msun$,
  $\alpha=1.2$ for $0.5 < m/\msun < 1$ and $\alpha=1.7$ for
  $m>1\,\msun$), and~\cite{chabrier03} (with a power-law slope of
$\alpha=1.3$ in the $m>1\,\msun$ regime, and $\phi(m) \propto
\exp(-(\log(m) -\log(m_c))^2/2\sigma^2)$ with $m_c=0.079$ and
$\sigma=0.69$ for $m<1\,\msun$).  In the top panel we show the four
functional forms as a function of mass and in the lower panel we
report the ratio with respect to the Salpeter IMF. All the reported IMFs
are normalized between $0.1$ and $100\,\msun$.

The universality of the IMF is actually still matter of
debate~\cite[see][for a review on IMF variations]{bastian2010}, with a
number of works exploring the possibility of an IMF functional form
that changes with galaxy environment and morphology~\cite[see,
  e.g.][and references therein]{bernardi2018}, as well as with time.

\subsection{Embedding chemical models into hydrodynamical simulations}
\label{sec:chem-sims}
Cosmological hydrodynamical simulations including the treatment of
chemistry have been developed by several groups in the last couple of
decades, starting with the early works by~\cite{raiteri1996},
\cite{lia2002} and ~\cite{kawata2003} on galaxies, comprising
Milky-Way-like objects and ellipticals, also in a cosmological context
as in~\cite{steinmetz1994} and~\cite{mosconi2001}.

Simulations of galaxy clusters and larger cosmological volumes also
started to include models of chemical evolution since the early 2000s.
\cite{valdarnini2003} presented results on the iron profiles of a set
of galaxy clusters obtained with an SPH set of cosmological chemical
hydrodynamical simulations.  In that work, the enrichment from SNIa
and SNII was considered, together with a metallicity-dependent cooling
function. In the paper the author discusses the effects on the ICM
iron profiles depending on the numerical resolution and on model
parameters (IMF, metal ejection profile and diffusion length).
Later, the first
implementations of chemical evolution modules were included into the
GADGET-2 code~\cite[][]{springel2005}, that is still one of the most
commonly used SPH codes for cosmological
simulations. These implementations only included metal production from SNII
under the assumption of instantaneous recycling approximation
and no metallicity dependence of the cooling function.
\cite{scannapieco2005} presented results on the enrichment
in simulated galaxies by adopting a chemical model in GADGET-2 that
included star formation, cooling, feedback and metal enrichment from
SNIa and SNII.  \cite{tornatore2004,tornatore2007} presented a more
detailed implementation of the chemical evolution model into GADGET-2,
which comprises a refined treatment for metallicity-dependent
cooling, metallicity-dependent yields and included the enrichment due
to intermediate- and low-mass stars undergoing the AGB phase~\cite[see
  also][]{dave2008}.
\cite{wiersma_chem2009,wiersma2009} also presented chemical
hydrodynamical simulations performed with the SPH code GADGET-2, where
they instead adopted metallicity-dependent lifetimes
by~\cite{portinari1998}.
In the 2000s, other groups working with different SPH codes also
started to include detailed models for chemical evolution in
simulations of galaxy clusters.  For instance, \cite{romeo2006}
presented zoomed re-simulations of galaxy clusters performed with the
TREESPH code by \cite{sommerlarsen2003}, including chemical evolution
with non-instantaneous recycling of gas and heavy elements (based on
the model by~\cite{lia2002}), in addition to metal-dependent cooling,
star formation, feedback from starburst-driven galactic winds and
thermal conduction.

More recently, advanced cosmological hydrodynamical simulations
performed with various codes, both SPH and grid-based, have combined
chemical models with several other physical processes, including
feedback from stellar and active galactic nuclei (AGN) sources, with
the purpose of obtaining more realistic simulations of galaxy clusters
that can be compared against the increasingly detailed observational
databases
available~\cite[][]{sijacki2006,puchwein2008,fabjan2010,teyssier2011,dubois2011,planelles2014,martizzi2016,biffi2017,dolag2017,barnes2017,vogelsberger2018}.

In addition to computing the global metallicity in the gas and stellar
components of the simulations, many of the modern chemical
hydrodynamical codes are able to trace individual metal species as
well, including not only iron, but also several additional chemical
elements, such as oxygen, silicon, calcium, nickel, etc.  Moreover, in
some simulations, the source of the enrichment is also specifically
tracked, which allows to reconstruct how much of the metal budget is
produced by each one of the followed enrichment channels, namely SNIa,
SNII and
AGB~\cite[][]{tornatore2007,biffi2017,biffi2018,vogelsberger2018,naiman2018}.

\subsubsection{Impact of the model assumptions}

When dealing with chemical hydrodynamical simulations, there are a
number of crucial aspects to take into account. In particular,
the choice of the IMF, of the stellar lifetime function and of the set
of stellar yields
have an ultimate influence on the resulting metal enrichment.
The amount of metals, in turn, is crucial because has also an impact
  on the cooling of the gas --- which has a strong dependence on
  metallicity~\cite[e.g.][]{sutherland1993,maio2007,wiersma2009} ---
  and consequently on the history of star formation.

\begin{figure}
  \centering
  \includegraphics[width=0.55\textwidth]{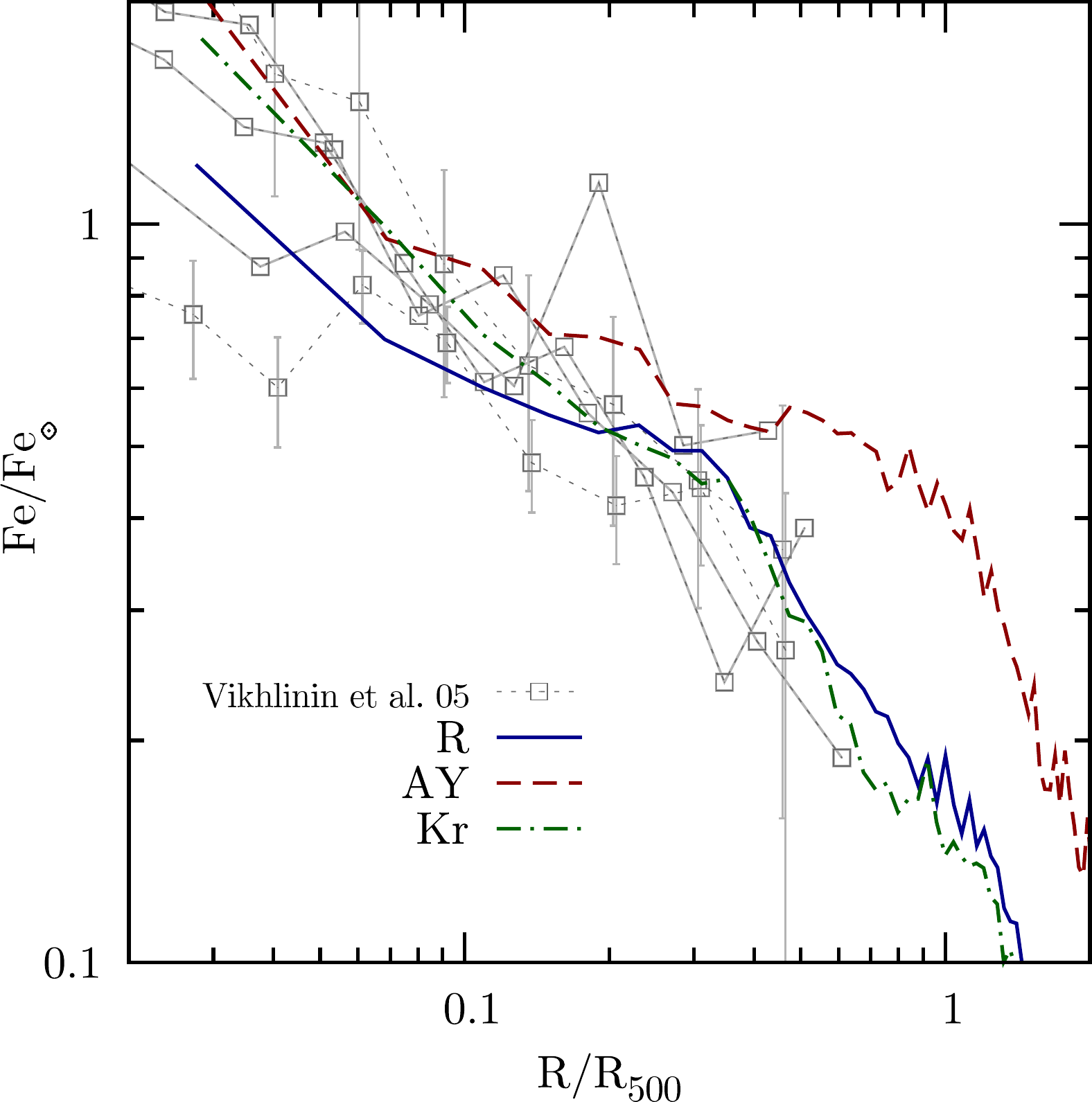}
  \caption{Effects of changing the IMF on the resulting Fe
    abundance profile; taken from~\protect\cite{tornatore2007}.
    Simulated Fe profiles assuming the IMF by Salpeter (R; blue solid line),
    Kroupa (Kr; green dot-dashed line)
    and~\cite{arimoto1987} (AY; red dashed line) are compared
    against observational data by~\protect\cite{vikhlinin2005} (grey simbols).
    Iron abundances are rescaled with respect to the reference solar values
    \mbox{by~\protect\cite{angr1989}.}}\label{fig:imf-effects}
\end{figure}

The effects of varying the assumed IMF or the set of stellar yields
can be significant.  As we mentioned in the previous
sections, different IMFs provide different relative ratios of
low-mass and high-mass stars, which reflects into different abundance
ratios.
For instance, \cite{romeo2006} presented an analysis on the effects of
varying the IMF on the ICM enrichment in simulations of galaxy
clusters, showing that in simulations where an efficient feedback
mechanism is not present and iron abundance profiles result to be too
steep compared to observed ones, adopting a top-heavier IMF rather
than a Salpeter IMF can alleviate the discrepancy. Similarly,
\cite{tornatore2007} found that changing the IMF, among the various
model assumptions, has the strongest impact on the resulting metal
enrichment of the ICM. In particular they
showed that assuming a top-heavier
IMF~\cite[such as the one proposed by][]{arimoto1987} with respect to
the Salpeter one turns into an increase of the ICM iron metallicity by
a factor of two and that of oxygen by a factor of three, with a
consequent increase of the oxygen-over-iron abundance ratio.
The authors also showed that the assumption of the multi-slope
Kroupa IMF~\cite[][]{kroupa2001}, on the
contrary, results into a decrease of the O/Fe abundance ratio with
respect to a Salpeter IMF, due to a smaller number of massive stars.
In this case, however, the Fe abundance profile is very similar to the
Salpeter one, at large cluster-centric distances ($r\gtrsim 0.2\,R_{500}$).
These results by~\cite{tornatore2007} on the Fe radial profile, for
the three simulation runs obtained for the IMF by~\cite{salpeter1955},
\cite{kroupa2001} and~\cite{arimoto1987}, are reported in
Fig.~\ref{fig:imf-effects}.

Given the sensitivity of element abundance ratios to the relative
distribution of different types of SNe and AGB stars, not only
  the proportion given by the IMF is important but also the different
time scales
contribute to different signatures in the final enrichment.
The most important effect of changing the lifetime functions,
  however, is on the enrichment history, post-poning or anticipating
  the typical epoch of ICM enrichment.

The amount of metals produced, and the relative abundances of
different elements, in simulations also depends directly on the set of
stellar yields adopted.  Given the uncertainties thereto related,
simulation analyses that investigate the chemical content of clusters
and galaxies should always clearly specify the sets of yields used,
given the uncertainties thereto related.
In fact, as shown by~\cite{wiersma_chem2009}, the predictions of
different sets of yields can differ by factors of a few in the ejected
masses of individual elements released by AGB and SNII, which
translates into uncertainties by a factor of a few in the elemental
ratios, for a fixed IMF.
The effects of these differences onto the final ICM enrichment is
however not straightforward, given that the metal production depends
in a non-trivial way on the SSP characteristics, such as its initial
metallicity, which depends in turn on several factors (epoch of formation, star formation history etc.).
For instance, \cite{tornatore2007} showed that using the sets of SNII
yields by either~\cite{WW1995} or~\cite{chieffi2004}, despite their
different element production, produce relatively similar
Fe abundance and O/Fe ratio profiles.

Degeneracies in the model assumptions can alone alter the ICM final
chemical pattern, especially --- but not only --- in the level of the
enrichment.  Because of these degeneracies, the choice of IMF,
lifetime function and yield tables is therefore a delicate aspect.
In~\cite{vogelsberger2018}, for instance, the authors describe how
changing the details of their chemical evolution model from Illustris
to IllustrisTNG, with significant updates of the SNII yield tables and
of the SNIa rates, can ultimately cause sizeable differences in the
integrated metal production over one Hubble time, namely by up to a
factor of $\sim 2$ in the case of iron.  Nonetheless, we also remind
that other crucial physical processes, above all energy feedback,
interplay in a complex way with the chemical model, and simulations
must at the same time reproduce both thermo- and chemo-dynamical
properties of the ICM, and possibly properties of the cluster galaxy
population, in order to be successful.

\subsubsection{Numerical limitations}

Also the details of the chemical model implemented included in the
simulations and the numerical resolution can impact the resulting
chemical enrichment of the~ICM.

Unfortunately, an accurate treatment of gas mixing within large-scale simulations is challenging, given the difficulty in following the development of turbulence and mixing even when the small scales involved are resolved.
Eulerian hydrodynamical codes implicitly include some numerical
diffusion, which can produce mixing that depends on the
specific implementation adopted and is often strong~\cite[e.g.][]{springel2010}.
Differently, Lagrangian methods such as SPH require the implementation of
an explicit mixing scheme, for instance adopting a diffusion model.
\cite{greif2009} proposed a method of chemical mixing in SPH simulations
that is based on the velocity dispersion within the SPH smoothing kernel.
Although applied to the evolution of a supernova remnant, this algorithm
is quite generic and can be suitable for simulations of the chemical enrichment of the interstellar and intergalactic medium.
Recently, \cite{williamson2016} presented a series of SPH simulations of
isolated galaxies performed with the $N$-body/smoothed-particle-chemodynamics
code GCD+~\cite[][and references therein]{kawata2014}, including also
chemical enrichment with a metal diffusion algorithm based on the method by~\cite{greif2009}. By comparing different metal mixing schemes,
the authors show that, depending on the diffusion strength,
the metals can flow out of the galactic disk without any resolved
mass flow or rather be retained and transported by galactic outflows.
This has therefore a critical impact on determining the dominant processes
in place in a galaxy, and can in turn influence their chemical feedback
to the surrounding medium.
Despite the few attempts done so far to include the treatment of diffusion processes
in SPH simulations, we remind that these have been typically limited to the scale of
the smoothing kernel. Nonetheless, transport mechanisms on larger scales might also
play an important role and much more realistic and detailed analyses
are still strongly~needed.

A frequently adopted approach to modelling chemical mixing in SPH codes,
especially for cosmological large-scale simulations,
is to assume that the products of
stellar nucleosynthesis are distributed within a fixed volume.
The distribution of metals from the stellar source onto
the surrounding gas elements is typically done by smoothing the
metallicities onto neighbour gas particles according to the same SPH
kernel used for the computation of the hydrodynamic
quantities~\cite[e.g.][]{mosconi2001}.  Tests presented
by~\cite{tornatore2007} showed that changing the kind of filter function,
the number of neighbour gas particles or the weighting scheme used to
perform the smoothing has a modest impact on the overall stellar
population and ICM enrichment in simulated clusters.
The differences introduced by these choices are typically smaller than the cluster-to-cluster variations.
\cite{valdarnini2003} also explored the variations on the ICM iron
profiles due to changes in the model parameters of the metal ejection scheme,
for a given IMF, and found changes by a factor of 1--2 in the normalization of the resulting iron radial distribution.

Another crucial aspect is the impact of numerical resolution
on the predictions from simulations on the ICM enrichment,
as discussed in several works in the
literature~\cite[e.g.][]{valdarnini2003,tornatore2007,wiersma2009,martizzi2016,vogelsberger2018}.
In the early investigations by~\cite{valdarnini2003}, cluster
simulations with increased numerical resolution provided generally
consistent iron abundance profiles with respect to the low-resolution
reference runs.

\begin{figure}
  \centering
  \includegraphics[width=0.85\textwidth]{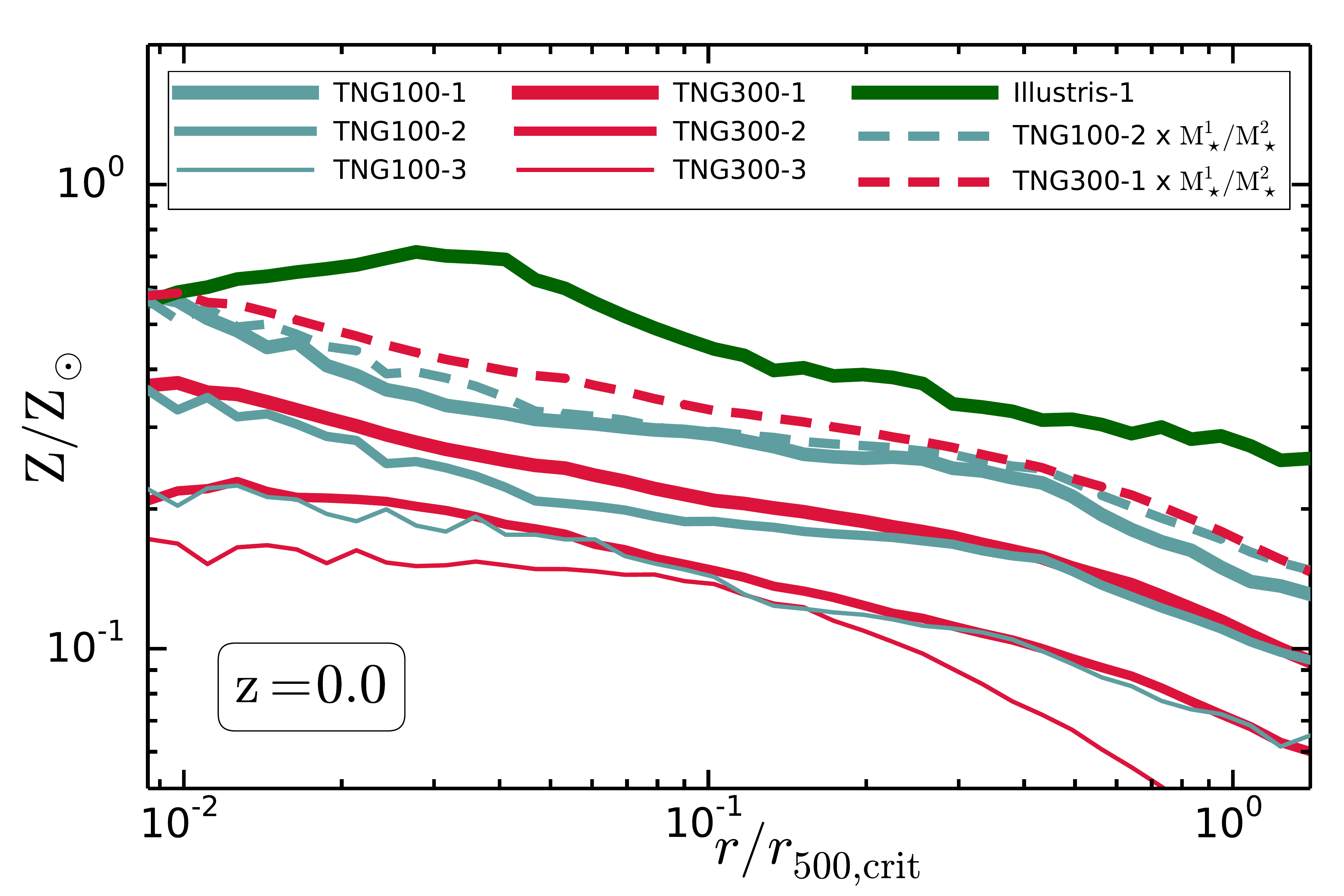}
  \caption{Resolution effects on the median
    (total) metallicity profiles for simulated clusters extracted from
    the Illustris and IllustrisTNG simulations; taken
    from~\protect\cite{vogelsberger2018}.
    Different lines refer to different resolution runs, as in the legend.
    The total metallicity is
    expressed in units of the solar value.}\label{fig:res-effects}
\end{figure}

Later numerical analyses indicated instead as a general finding that the level of enrichment of the ICM tends to increase at higher resolution. \cite{tornatore2007}, for instance, show that the effect of resolution is modest, albeit sizable, at redshift $z=0$ and more prominent at higher redshift, with differences in the enrichment history by a factor of $\sim 2$.
Similarly, \cite{vogelsberger2018} find that the metallicity profiles of the IllustrisTNG simulated clusters vary by a factor of $\sim 2$ with resolution, as reported in Fig.~\ref{fig:res-effects}.
Indeed, increasing the resolution effectively
enhances the star formation, especially at high redshift, as a
consequence of the larger number of resolved small haloes that
collapse first and in which gas cooling happens very efficiently.
Since this effect is more prominent at high redshift, the
pre-enrichment history in the simulations will change, resulting in a more efficient metal enrichment of the gas in the early epochs, that propagates till lower redshifts.
As explicitly discussed by~\cite{martizzi2016}
and~\cite{vogelsberger2018}, numerical convergence is however rather
slow and difficult to achieve. Furthermore, when numerical resolution
is increased the sub-grid physical models in the simulation need
to be re-calibrated to maintain the agreement with some reference observables,
and this does not imply convergence of results in the enrichment properties.

Finally, the physical processes treated in the simulations,
and their sub-grid implementation, also have a sizable impact on
the final enrichment history and metal distribution in
simulated clusters.
In the following sections, we will outline the impact
of the energy feedback scheme adopted on the distribution of
metals in the ICM, in comparison to observations.

\section{Abundance distribution in the ICM}\label{sec:3}

A number of physical and dynamical processes can influence the
distribution of metals in the ICM: galactic winds and AGN feedback,
depletion of metal-rich gas due to cooling and consequent star formation
or to dust formation, or dynamical processes like merging events,
large-scale motions and ram-pressure stripping~\cite[see][for a review]{schindler2008}.
Therefore, the study of chemical abundances in the hot ICM provides invaluable
information about the formation and assembly histories of the clusters
as well as on the interplay between the galactic population and the surrounding
diffuse medium.
Also, the spatial distribution of specific heavy ions in the ICM,
that are produced from different stellar sources on different
time scales, carries the imprint of the star formation and
assembly history of the cluster as well as the signatures of the
feedback processes from stars and AGN.
The comparison between the distribution of
different elements and their abundance ratios can unveil precious
information on the relative incidence of the stellar sources that
released them.

In this section we review the main results on the ICM
enrichment in cluster simulations with special attention to the
spatial distribution of heavy elements, from the innermost regions out
to the cluster boundaries.

\subsubsection*{Metallicity estimates in simulations}

Similarly to many other thermodynamical quantities, also metal
abundances can be computed in different ways from the intrinsic,
three-dimensional properties characterizing the hot gas component in
simulated clusters.
This aspect is particularly important when comparing against
observational datasets.
In fact, measurements of chemical abundances in the ICM are typically
derived from X-ray observations of the ICM spectral emission, which is
essentially composed by a Bremsstrahlung continuum and emission lines
of heavy ions~\cite[][\Mrev]{werner2008,boehringer2010}.
This measurement is therefore naturally sensitive to the
  emissivity of the gas.

In simulations, the value of metallicity for the gas contained in a given region is typically computed as
\be
Z_w = \frac{\int w Z {\rm d}V}{\int w {\rm d}V}
\to
\frac{\sum_i Z_i w_i}{\sum_i w_i}\, , \label{eq:Zw}
\ee
\looseness=-1 where the second fraction is the discretized formula for
the gas elements considered in the simulations. Specifically,
$Z_i$ and $w_i$ are the metallicity and
chosen weight of each gas element $i$ selected. In general, two weighted averages are most commonly
adopted, the {\it mass}-weighted value and the {\it emission}-weighted one. In the former the weight is simply the mass of the gas element,
$w_i = m_i$. In the latter, the weight is the gas emissivity $w_i =
m_i \rho_i \Lambda(T_i,Z_i)$, where $m_i$ and $\rho_i$ are the
gas-element mass and density, and $\Lambda(T_i,Z_i)$ is the cooling
function, which depends on the temperature ($T_i$) and global
metallicity of the gas element.
Depending on the definition adopted, the metallicity of the single gas element ($Z_i$) is given by the fraction between the metal mass and the gas or hydrogen mass associated to it.
Similarly, in simulations where the single metal species are separately tracked, the weighted average expressed by eq.~(\ref{eq:Zw}) can be calculated for individual metal abundances~($Z_i^X$, for the chemical element $X$).

For a more faithful comparison with X-ray data, one could generate synthetic X-ray observations out of the simulation outputs.
To this scope, a number of
groups have dedicated a significant effort, designing specifically
suited X-ray simulators able to generate mock X-ray data mimicking the
performances of existent and upcoming X-ray
telescopes~\cite[][]{xmas2004,xim2009,phox2012,pyPHOX2014}.
Using this technique, \cite{rasia2008} investigated possible
observational biases in the X-ray derived metallicity measurements
from simulated clusters by analysing mock \XMM\ spectra. Their main
findings are that iron abundances are well recovered with respect to
the intrinsic simulation value whenever the gas plasma has a
temperature higher than 3\,keV or lower than 2\,keV, while the
multi-temperature structure of gas in the 2--3\,keV temperature range
typically leads to an overestimate in the spectroscopic
measurement. Oxygen, on the contrary, is overall well measured for
low-temperature ($T< 3$\,keV) systems while it can be overestimated
for hotter clusters.
We note that the test presented in~\cite{rasia2008} was specifically
done for CCDs and that oxygen, in particular, is measured at the
limits of the energy band sensitivity.  A similar approach has been
adopted by~\cite{cucchetti2018}, to predict the capabilities of the
X-ray Integral Field Unit (X-IFU) on board the next-generation
European X-ray observatory \Athena\ in reconstructing the chemical
properties of the ICM from mock observations of a sample of simulated
galaxy clusters. In this work the authors show that
emission-measure-weighted metallicity values (obtained
from~\eqref{eq:Zw} with $w_i = m_i \rho_i $) computed from the
simulations can be accurately recovered from the spectral analysis of
the mock X-ray images up to redshift $z\approx 2$.

\subsection{Global distribution}
\label{sec:glob}

From the global distribution of particular elements in the ICM
valuable information can be inferred on the underlying distribution of different types of supernovae or on the feedback processes in the ICM.

The influence of varying the IMF and therefore the relative number of SNIa and SNII is in fact visible from the metallicity distributions shown in~\cite{tornatore2007}. Typically, they find that SNII enrichment (e.g., oxygen) dominates patchy high-density regions located in the vicinity of star-formation sites, while AGB and SNIa contribute more to the diffuse enrichment. The contribution of SNII is more evident when a top-heavier IMF is assumed, which enhances the relative number of massive SNII-progenitor stars~\cite[see also][]{fabjan2008}. These simulations accounted only for stellar feedback, but changing the details of this feedback scheme included also has an impact on the global distribution.
In particular, \cite{tornatore2007} show that increasing the winds strength has the general effect of increasing the diffusion of metals. This effect is more significant for oxygen than for iron, because winds are uploaded with star-forming gas particles and therefore mostly enriched by~SNII.

\begin{figure}
  \centering
  \includegraphics[width=0.95\textwidth]{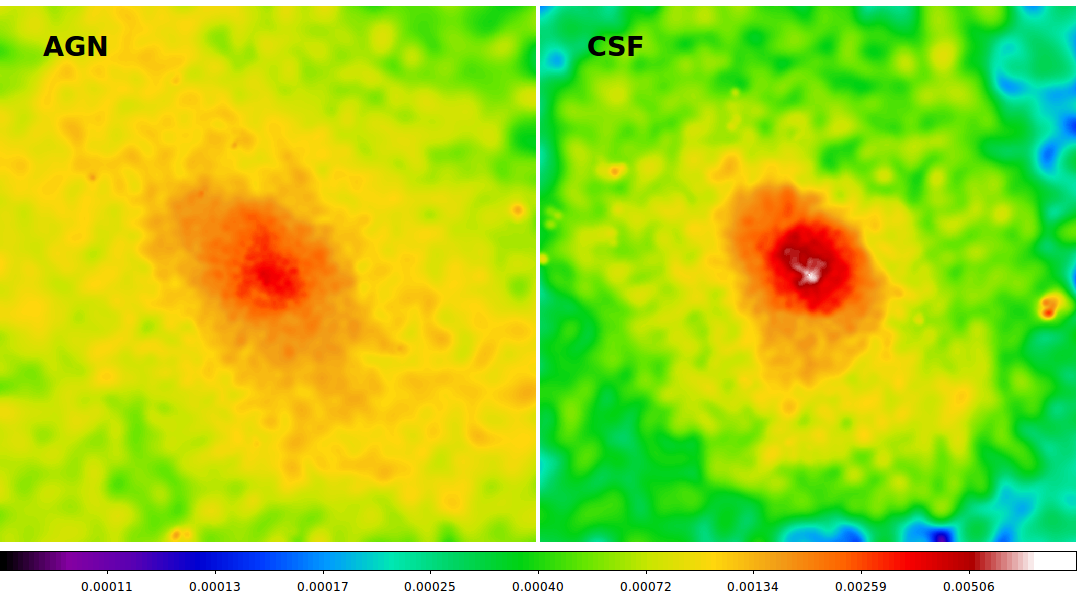}\\
  $Z_{\rm O}$
  \caption{Mass-weighted oxygen abundance maps for a cluster at $z=0$,
    simulated with stellar and AGN feedback (`AGN'; left) and
      only with stellar feedback (`CSF'; right).  The oxygen
    abundance, $Z_{\rm O}$, is the mass fraction w.r.t. to
    hydrogen. Each map has a side of 2\,Mpc, is projected for 2\,Mpc
    along the line of sight, and is centered on the minimum of the
    cluster potential well. The cluster belongs to the simulation set
    for which results on the ICM metal enrichment have been presented
    in~\cite{biffi2017,biffi2018}.\label{fig:maps}}
\end{figure}
From Fig.~\ref{fig:maps} we can see how the global distribution of
oxygen (mass-weighted) abundance in the ICM changes depending on the
feedback scheme included in the simulations. The two clusters
reported~\cite[see][for more details on the simulations]{rasia2015}
are the same object, simulated with different feedback schemes: on the
left, stellar and AGN feedback are both included (`AGN' simulation
run), while on the right only the former is considered (`CSF'
simulation run).  Visibly, the present-epoch level of oxygen
enrichment and its spatial distribution are very different. In the
`AGN' case the gradient with cluster-centric distance is flatter and
more homogeneous especially at large radii with respect to the `CSF'
map, where the distribution is clumpier, more centrally-peaked and
with more peaked sub-structures (e.g. the one in the right-most
periphery)~\cite[see also][]{fabjan2010}.  Given that oxygen is
released by SNII on short time scales and therefore traces recent star
formation, the more homogeneous distribution in the `AGN' cluster is
an evidence of pre-enrichment. Indeed, AGN feedback at early epochs is
able to widely distribute metal-enriched gas outside small haloes
already before the star formation
peak~\cite[see][]{biffi2017,biffi2018}.
This effect is ultimately reflected in the present-day radial profiles of clusters, especially in the outermost regions (see further discussion in Sec.~\ref{sec:outsk}).

\subsection{Radial metallicity profiles}
\label{sec:rprof}

As inferred from the maps in Fig.~\ref{fig:maps}, the study of the
metallicity profiles in clusters and of their gradient provides direct
information on the efficiency of the feedback history of galaxy clusters.

Observationally, radial metallicity profiles of the ICM have been extensively investigated since the first explorations of bright nearby clusters with \ASCA\ and \Beppo. The most recent X-ray observatories, namely \Chandra, \XMM\ and \Suzaku, were able to map the ICM enrichment from the innermost region out to the cluster boundaries (see \Mrev).  Already from the first observational evidences, it has been clear that
ICM metallicity profiles present gradients
with differences in the central slope depending on the cool-coreness of the cluster.

In cosmological hydrodynamical simulations of galaxy clusters radial
distribution of the total ICM metallicity or of specific element
abundances (when individually traced) have been investigated at
various levels of complexity, in terms of the physical processes
treated within the code.
Many numerical works that include only stellar feedback typically
found iron abundance profiles steeper than the observed profiles. For
instance, \cite{valdarnini2003} analysed a set of simulations of
regular galaxy clusters characterized by a quiet dynamical state,
following the production of iron from SNIa and SNII, and found that
the simulated iron profiles showed a radial decrease in the outer
regions that was clearly steeper than the observational profiles
by~\cite{degrandi2001}. This steeper gradients of the metallicity
profiles have been connected to a lack of an efficient feedback, such
as the one associated to
AGN~\cite[][]{sijacki2006,fabjan2010,planelles2014,biffi2017}.\footnote{The
  different slope of the metallicity profiles in simulations with and
  without AGN feedback can be seen from the comparison in
  Fig.~\ref{fig:out}.}

\cite{tornatore2007}, despite including only stellar feedback, found a
better agreement with observed profiles by~\cite{vikhlinin2005}, but
showed how the details of the feedback model (e.g.\ the velocity of
the winds) affect the final shape of the iron abundance profile. As
remarked by the authors, not only the ICM metallicity profiles but also the properties of the
galaxy population must be successfully reproduced. However, simulations
based on stellar feedback only typically produce central
cluster galaxies that are bluer and more star forming than observed.
A recent investigation by~\cite{martizzi2016}, including both stellar
and AGN feedback, also showed metallicity gradients in clusters
simulated with an AMR code that are fairly consistent with the
observational data by~\cite{leccardi2008} and~\cite{matsushita2013},
despite a disagreement in the normalization. A flatter shape towards
the outskirts is also recovered, when the emission-weighted
metallicities are computed.
A better agreement with recent observed profiles, also in term of
normalization, was instead shown for the ICM iron profiles in
simulated galaxy clusters by~\cite{biffi2017,biffi2018}
and~\cite{vogelsberger2018}, with different numerical codes,
resolution and implementation of the main feedback processes.
We will discuss more later in Section~\ref{sec:outsk} how the effect
of early AGN feedback in numerical simulations of clusters helped
reproducing the flat outskirt profiles found by X-ray observations.

\begin{figure}
  \centering
  \includegraphics[width=0.6\textwidth]{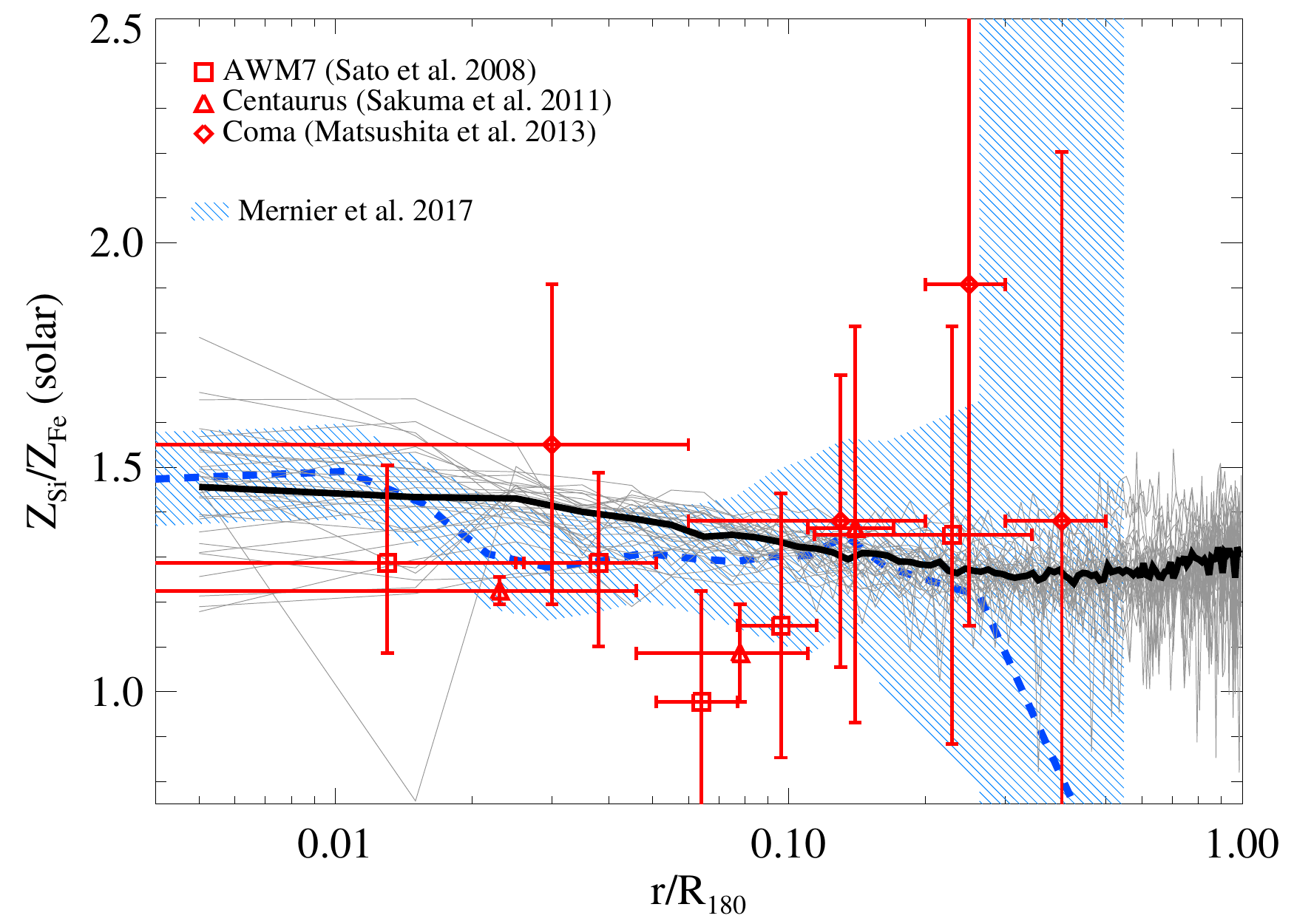}
  \includegraphics[width=0.6\textwidth]{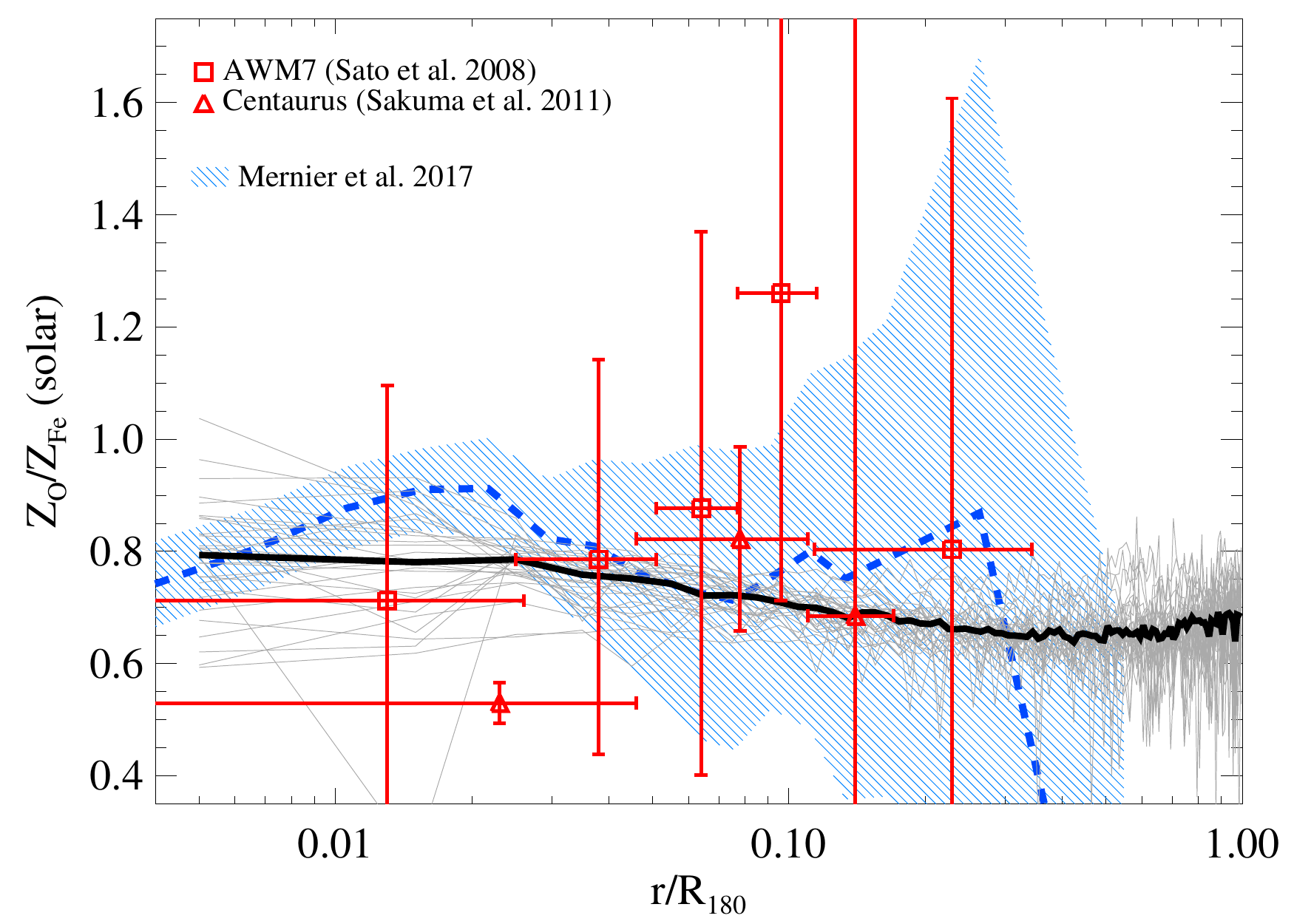}
\caption{Median abundance ratio profiles as a function of cluster-centric distance for the set of simulated clusters presented in~\protect\cite{biffi2017} (black solid line; individual profiles are marked as gray thin lines). {\it Upper panel:} Si/Fe; {\it lower panel:} O/Fe. For comparison, observational data by~\protect\cite{mernier2017} (blue shaded areas and dashed lines), \protect\cite{sato2008} (red squares), \protect\cite{sakuma2011} (red triangles) and \protect\cite{matsushita2013b} (red diamonds) are reported as well; all abundances are rescaled with respect to the reference solar values by~\protect\cite{angr1989}.
  \label{fig:profs}}
\end{figure}

Other than iron or total metallicity, the distribution of other
elements such as oxygen or silicon, and their relative abundance with
respect to iron, provide valuable information on the distribution of
SNIa and SNII.  Despite previous findings indicating an excess of Fe
with respect to SNII products in the center, many recent observations
suggest that abundance ratios are found to be remarkably uniform from
the core out to large cluster-centric
distances~\cite[e.g.][\Mrev]{sato2008,sakuma2011,matsushita2013b,simionescu2015,mernier2017}.
From the simulation side, recent results by~\cite{biffi2017} show
similarly flat abundance ratio profiles compared to X-ray data for the
Coma, Centaurus and AWM7 galaxy
clusters~\cite[][]{sato2008,sakuma2011,matsushita2013b}, with a mild
increasing trend towards the center~\cite[see
  also][]{planelles2014}.
These results are reported in Fig.~\ref{fig:profs}, for the Si/Fe
(upper panel) and O/Fe (lower panel) abundance ratios. The median
abundance ratio profiles from the simulated clusters by
\cite{biffi2017} are compared with the X-ray data for the Coma,
Centaurus and AWM7 clusters and with the mean profiles reported
by~\cite{mernier2017} for a sample of 44 nearby cool-core galaxy
clusters, groups, and ellipticals~\cite[from the CHEERS
  sample;][]{mernier2016,deplaa2017}.  From the comparison between
simulated and observed profiles, we note a remarkable agreement in the
intermediate/outer radial range for both abundance ratios, especially
considering the scatter around the median profiles. Interestingly,
this result holds not only for the trend but also for the
normalization, despite the possible uncertainties due to the set of
stellar yields adopted in the model. In the Si/Fe case, in particular,
the median simulated profile is in very good agreement with
observational data also in the central region.
Instead, a mild discrepancy is observed at small radii, $r\lesssim
0.04\,R_{180}$, between the median simulated O/Fe profile and the
  one by~\cite{mernier2017}, although simulated and observed data are
still consistent within the scatter. We remind nevertheless that the
accurate measurement of oxygen abundances with CCD instruments is
still very challenging.

In this respect, also the Si/Fe and O/Fe profiles presented by~\cite{vogelsberger2018} are consistent with the observational values reported in Fig.~\ref{fig:profs}, given the large error bars in the observed profiles~\cite[see Fig.~7 --- r.h.s.\ panels --- in][]{vogelsberger2018}.
Nevertheless, the abundance ratios by~\cite{vogelsberger2018} present
an opposite tendency to slightly increase from the center towards the outskirts with respect to the flatter profiles by~\cite{mernier2017}.

In both these simulation sets the relative contribution from SNIa and
SNII to the ICM metal content as a function of radius has been
inspected directly, confirming that the bulk of the metal production
(in terms of mass fraction) at all radii is largely dominated by
SNII. Also, in both cases abundance ratio profiles, such as Si/Fe and
O/Fe, are very good tracers of the SN relative contributions and of
the different ICM enrichment histories, mirroring in fact in the
different trends with radius highlighted above.

\subsection{Cluster core}\label{sec:core}

Observations of strongly peaked X-ray emissivity in the centre of
relaxed clusters suggest the presence of significant cooling flows in
the central regions of galaxy clusters as a consequence of the
radiative losses of the ICM in the core~\cite[][]{fabian1994}, that
would become denser and colder initiating a runaway cooling process.
Even though observations show no presence of such massive cooling
flows, suggesting that a central source of heating (e.g. AGN feedback)
must be at play, the low-entropy dense gas does sink towards the core
of the so-called cool-core (CC) clusters~\cite[named as such
  by][]{molendipizzolato2001}.
Differently, non-cool-core (NCC) clusters are typically observed to have
nearly isentropic gas cores at a higher entropy level~\cite[e.g.][]{maughan2008}.
A definite distinction from CC, nonetheless, is still largely debated and the
definition of CC itself based on the central thermal properties (e.g.\ central density or temperature drop, central cooling time or mass deposition rate) can vary substantially from author to author~\cite[][and references therein]{barnes2018,hudson2010}.
Starting with \Beppo\ and \XMM\ observations of large cluster samples
in the early 2000s~\cite[e.g.,][]{degrandi2001,degrandi2004,leccardi2008}, X-ray
observations indicated that also the central gradient of the ICM metallicity
is different depending on the cluster cool-coreness (see also \Mrev).

\begin{figure}
  \centering
  \includegraphics[width=0.99\textwidth]{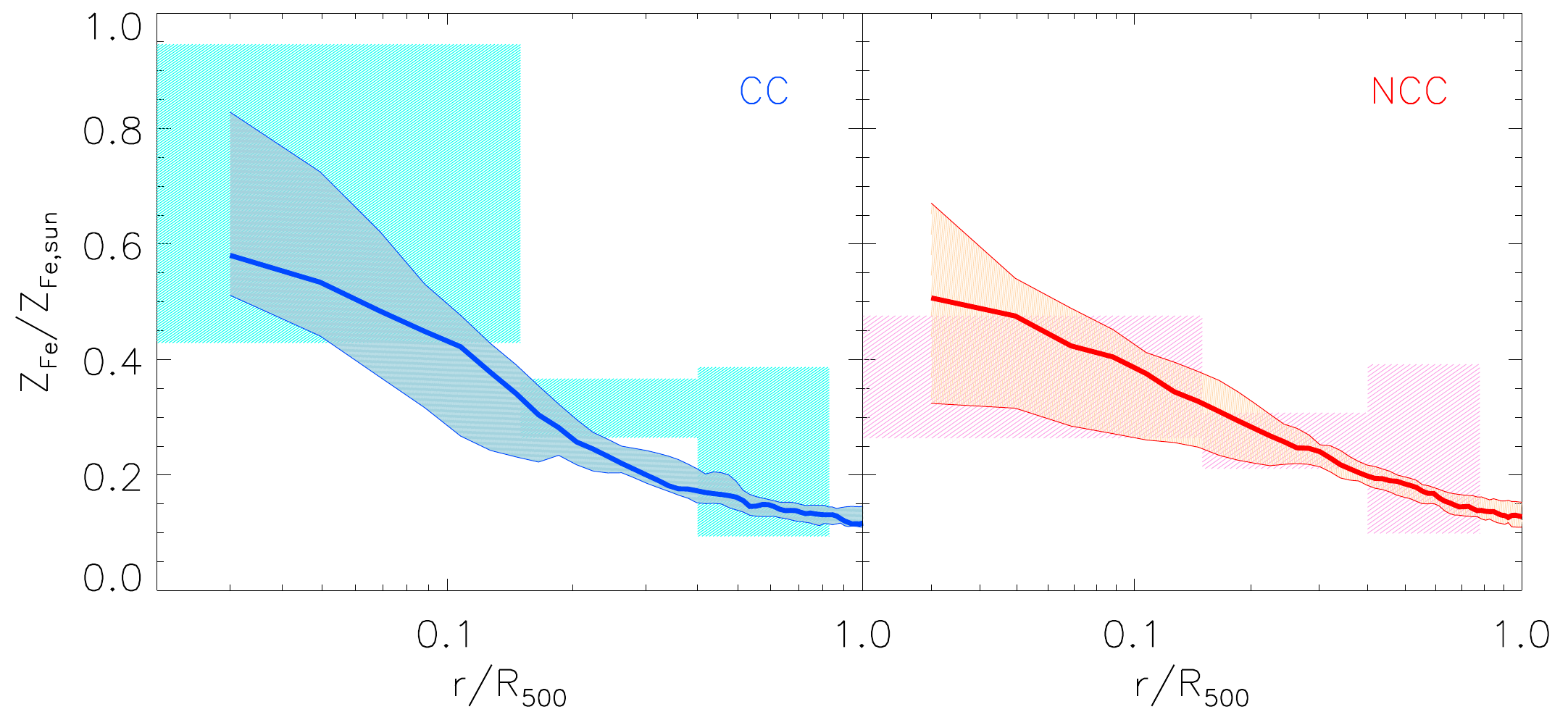}
  \caption{CC (left) and NCC (right) median metallicity profiles
    for the simulated clusters at $z=0$ presented
    in~\cite{rasia2015} (solid lines, with $16$th and $84$th percentiles around the median). Shaded boxy
    areas in the two panels represent observational data for local ($z<0.2$) clusters
    by~\protect\cite{ettori2015}. All
    abundances are rescaled with respect to the reference solar values
    by~\protect\cite{angr1989}.}
  \label{fig:profs-cc}
\end{figure}

From the theoretical point of view, cosmological hydrodynamical simulations
have struggled for many years to reproduce the observed diversity between CC and
NCC clusters:
simulated clusters were typically characterised by higher central metallicities
than observed, independently of the central entropy level,
that was generally high like in NCC systems~\cite[][]{dubois2011}.
Only recently, cosmological hydrodynamical simulations have been able
to recover the observed
diversity of CC and NCC populations~\cite[][]{rasia2015,hahn2015,barnes2018},
both in thermal and chemical properties.
In agreement with observations, these numerical investigations have found
that the metallicity profiles are flatter for NCC systems,
whereas they present a peak in the core of CC clusters, for which the metallicity
gradients result to be steeper~\cite[see also][]{martizzi2016,vogelsberger2018}.
This is shown in Fig.~\ref{fig:profs-cc}, where we report the median CC and NCC Fe profiles (l.h.s.\ and r.h.s.\ panel, respectively) for the sample of simulated clusters presented in~\cite{rasia2015} and~\cite{biffi2017}, compared to observations by~\cite{ettori2015}.

As shown in Fig.~\ref{fig:profs-cc}, while CC present a
metallicity peak towards the centre ($r < 0.15\,\rfive$), NCC show
instead a lower central enrichment level.
This chemical diversity anti-correlates therefore with the entropy level: CC are characterised by a decreasing entropy profile towards the center, with a core entropy lower than NCC systems, where, as previously said, the entropy profiles flattens to higher values.

\begin{figure}
  \centering
  \includegraphics[width=0.85\textwidth,trim=25 15 25 15,clip]{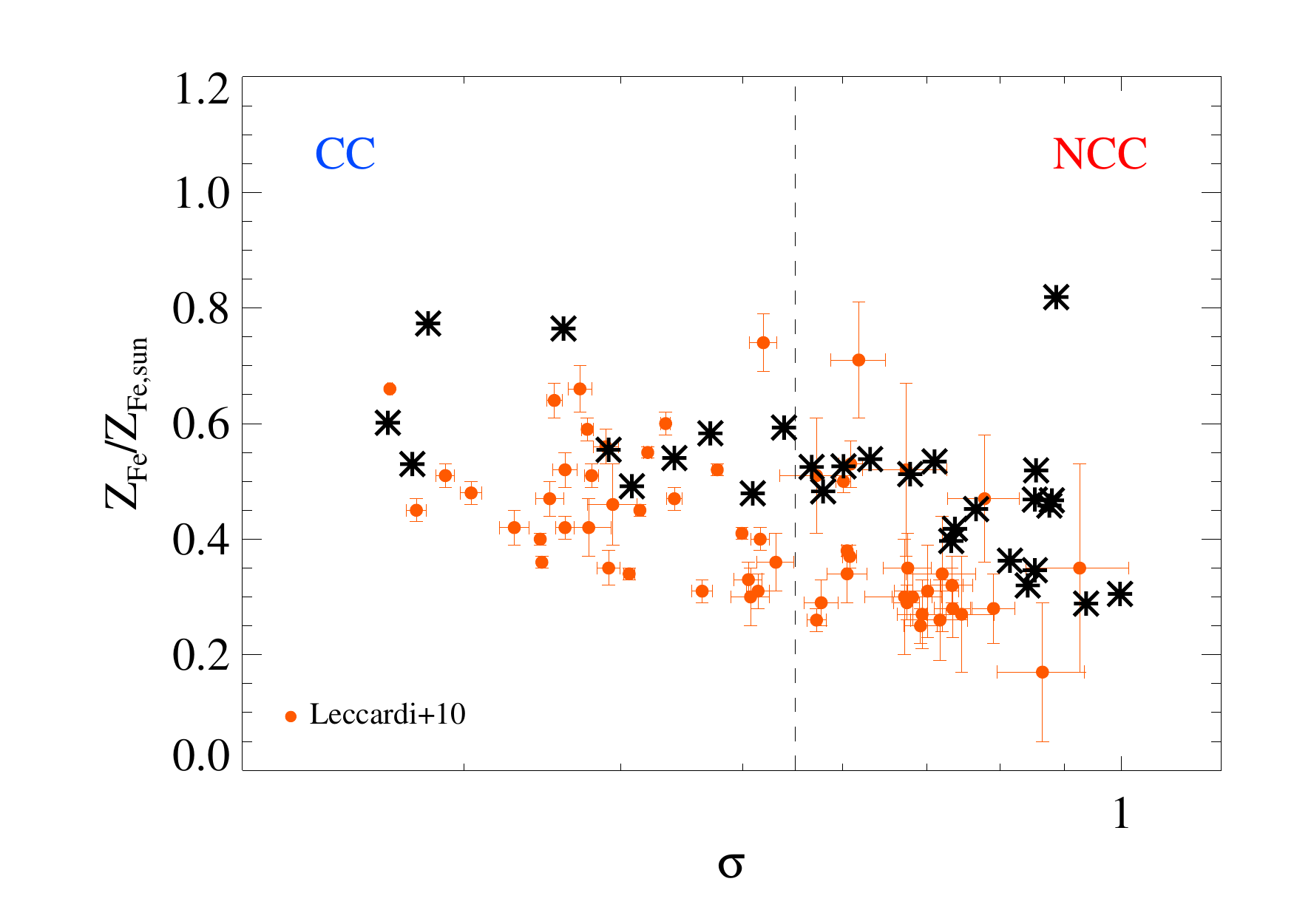}
  \caption{Relation between central metallicity and
    entropy for the simulated clusters from the sample presented
    in~\protect\cite{biffi2017} (black asterisks), compared to data
    by~\protect\cite{leccardi2010} (orange circles with error bars). All
    abundances are rescaled with respect to the reference solar values
    by~\protect\cite{angr1989}.
    \label{fig:core}}
\end{figure}

This anti-correlation between entropy and metallicity in the cores of galaxy clusters
is for example shown in \XMM\ observations by~\cite{leccardi2010}.
In Fig.~\ref{fig:core}, the relation between
core entropy and central Fe abundance is reported for the sample of simulated clusters analysed in~\cite{biffi2017} and compared to the
observational data by~\cite{leccardi2010}. Similarly to observations,
the central entropy gradient can be quantified by the pseudo-entropy
ratio $\sigma = (T_{\rm sl,IN}/T_{\rm sl,OUT}) * ({\rm EM}_{\rm IN}/{\rm EM}_{\rm OUT})^{-1/3}$, with the spectroscopic-like temperature~\cite[$T_{\rm sl}$;][]{mazzotta2004}
and emission measure (EM) computed within the projected ``IN''
($r<0.05\,R_{\rm 180}$) and ``OUT'' ($0.05\,R_{\rm 180} < r <
0.2\,R_{\rm 180}$) regions~\cite[e.g.][]{leccardi2010,rossetti2011}.
The Fe abundance is also computed within the
IN region
used to calculate $\sigma$.  The comparison shows that the two
quantities anti-correlate with a similar slope to observed one, with
CC clusters that are more metal-rich in the center with respect to NCC
clusters.
Essentially, this result indicates that the low-entropy gas is also
metal rich. For the dense low-entropy gas the radiative cooling is in
fact very efficient and star formation can take place, consequently
enriching the gas with metals. Given that diffusion of metals takes place
on long time-scales, the metals trace therefore the gas region where they
have been injected and can be diluted only if the ICM itself under-goes mixing processes.
Thus, since the low-entropy gas is also the one preferentially
sinking towards the center of the clusters, a low-entropy high-metallicity
core can form. In NCC systems, on the other hand,
the low-entropy metal-rich gas is likely to be more diffuse rather than
concentrated in the innermost region. A likely explanation for this is related to
the effect of merging events that displace and distribute the metal-rich low-entropy gas.

In~\cite{biffi2017} the authors also show the central entropy-metallicity relation
for clusters simulated without the inclusion of AGN feedback, finding that the majority of systems has high core entropy (like NCC) but also a central metallicity that is far larger than observed, making the entropy-metallicity relation too steep compared to data.

More generally, however, we note that differences between various
simulation results still persist in the modelling of cluster cores,
given that the detailed metal distribution in those regions is highly
sensitive to the details of the galaxy formation model, especially the
specific AGN feedback implementation.

\subsection{Cluster outskirts}
\label{sec:outsk}

The simulation results on cluster metallicity profiles (e.g. shown in
Fig.~\ref{fig:profs-cc})
are in line with X-ray observations not only in the central
regions but also in the outskirts, where little variance is found
between CC and NCC profiles.  In particular, X-ray observations
indicate a remarkably uniform ICM enrichment beyond the cluster core,
out to $\rfive$ or
more~\cite[][]{fujita2008,leccardi2008,werner2013,simionescu2015,urban2017}.

At intermediate and large cluster-centric distances the metallicity
and abundance ratio profiles are all found to be relatively flat~(see
\Mrev),
in both CC and NCC clusters~\cite[e.g.][as shown in
  Fig.~\ref{fig:profs-cc}]{ettori2015}.
\begin{figure}
  \centering
  \includegraphics[width=0.85\textwidth]{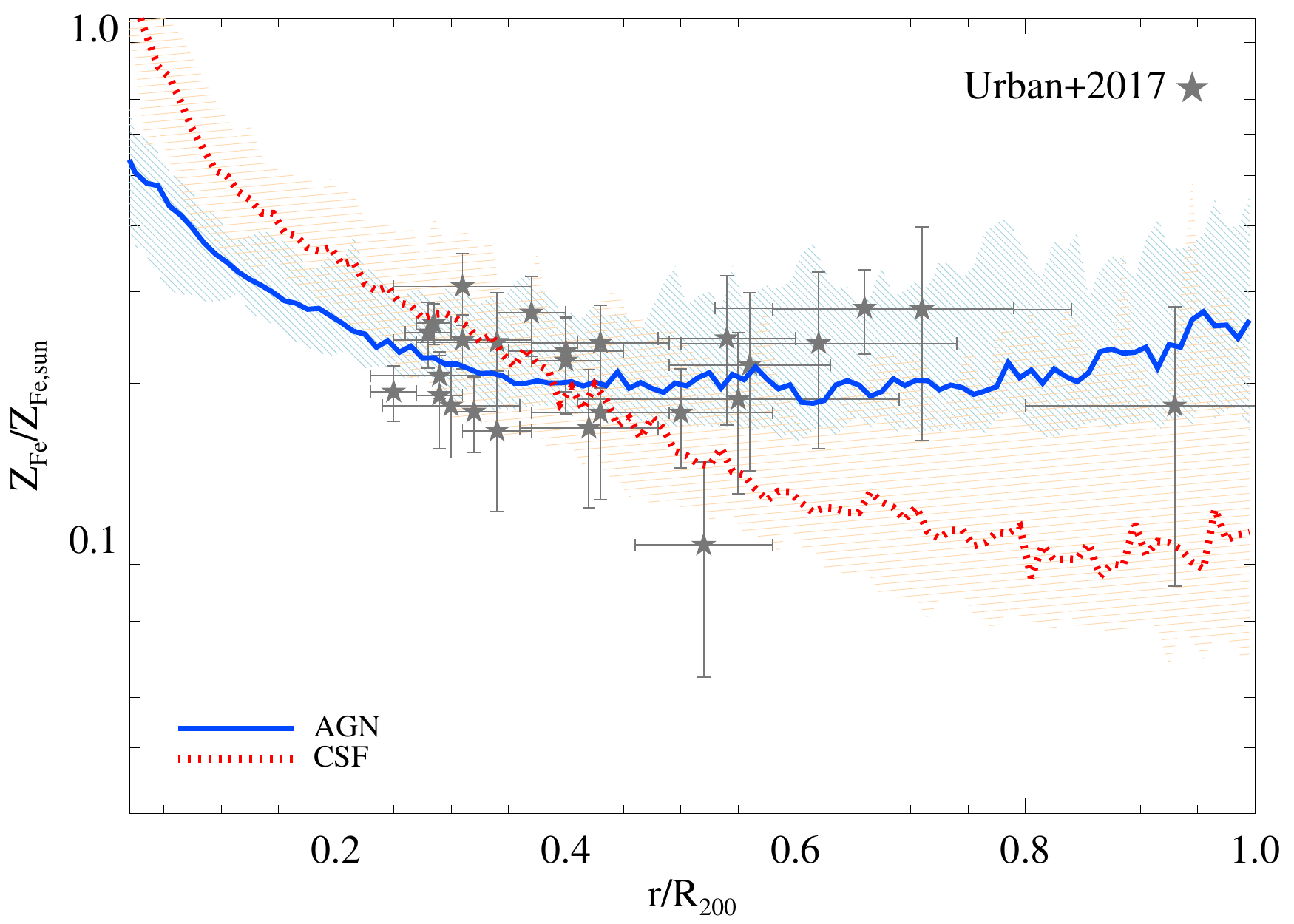}
  \caption{Comparison between Fe median profiles in the same
    simulations including only stellar feedback (`CSF') or both stellar
    and AGN feedback (`AGN'). The scatter on the profiles is marked by
    the shaded areas. For comparison, we also report data
    by~\protect\cite{urban2017} for the outskirts ($r > 0.2\,R_{200}$)
    of local galaxy clusters (symbols).  All abundances are rescaled
    with respect to the reference solar values
    by~\protect\cite{angr1989}.
    \label{fig:out}}
\end{figure}

The shape of the abundance profiles, instead of their amplitude, is a
key information to constrain the impact that different feedback
mechanisms have on the ICM enrichment pattern, since degeneracies
among the assumptions of the chemical model can play a role in the
normalization of the
profiles~\cite[e.g.][]{romeo2006,tornatore2007,fabjan2008}.
Early results from cosmological hydrodynamical simulations typically
found steeper profiles than observed, with a clear decreasing trend
with increasing radius.
Later on, many groups started to include more efficient energy
feedback processes, such as feedback from AGN, to alleviate the
central overcooling in groups and clusters~\cite[e.g.][and references
  therein]{borgani2011}, which eventually allowed to successfully
reproduce many realistic properties of galaxy clusters. Despite the
persisting challenges to reproduce correctly the core properties, the
role played by AGN feedback was found to be effective in producing
flatter ICM metallicity profiles at large radii $\gtrsim
0.2\mbox{--}0.3\,\rvir$~\cite[][]{sijacki2006,fabjan2010}.

This can be seen from the profiles reported in Fig.~\ref{fig:out},
based on simulation results presented in~\cite{biffi2017,biffi2018}.
In the figure, we show the median emission-weighted projected iron abundance
profiles~\cite[with respect to solar reference values by][]{angr1989}
for a sample of 29 simulated clusters. The simulations include
self-consistent treatments for gas cooling and star formation, metal
enrichment from stellar evolution, and energy feedback from both
stellar and AGN sources (`AGN' run; blue curve and shaded area). For
comparison, we also show the median profile for the same set of
clusters simulated without AGN feedback (`CSF' run; red curve
and shaded area). By inspecting the shape of the two profiles, it is
visible that the inclusion of AGN feedback in the simulations produces
an overall flatter profile than the `CSF' run, especially in the
outskirts ($r \gtrsim 0.2\,\rtwo$) where the former is in better
agreement with recent X-ray data from~\cite{urban2017}. Despite the scatter among different clusters present at large radial distances, the difference between the two runs is well defined: while the `CSF' profile
significantly declines, the `AGN' profiles remains flat, indicating a uniform enrichment in the outer regions.

The origin of this uniformity cannot be ascribed
to the ongoing activity of the central AGN
because not even an efficient AGN feedback can be able to displace significant
amount of metal-enriched gas from the central regions out to the
virial radius at low redshifts, especially in massive clusters where
the potential well is very deep. Thus, the uniform enrichment in the
outskirts of present-day massive galaxy clusters must be the result of
a pre-enrichment in which the role of early feedback processes appear to be crucial.
Recent observational results and their quantitative agreement with those from cosmological hydrodynamical simulations further support this picture.

Detailed investigations of the IllustrisTNG simulations
by~\cite{vogelsberger2018} on the origin of metal-enriched gas within
clusters at $z\sim0$ show that the majority of it was accreted from
the proto-cluster region. In particular, \cite{vogelsberger2018} show
that the average metallicity of the gas residing in the
$5$\,Mpc-radius around the cluster progenitors is already high at
$z\sim 2$~\cite[see also][]{biffi2017} and close to the typical
metallicity of the outer profiles, so that its later accretion onto
the cluster does not alter the typical outskirts enrichment and the
outer profiles remain nearly constant and flat down to $z\sim0$.  The
metallicity of the accreted gas is also quite universal across their
large cluster sample.
Similar results were found by~\cite{biffi2018}, who investigated the
origin of the metal-rich gas found in the outskirts (roughly
$\rfive<r<\rtwo$) of a study sample of four simulated clusters, selected to have different masses and core properties. By directly tracking back in
time and space the hot gas particles residing in the outskirts at
$z=0$ the authors showed that at higher redshift they were residing in the proto-cluster
region, mainly outside the progenitor, and are later
accreted onto the forming cluster.
At $z\sim 2$, the tracked gas comprises both pristine particles and
highly-enriched ones, which makes the average metallicity of this
accreted material already very similar to that in the outskirts.  In
this study, the authors also show that the relative contribution from
SNIa and SNII to the enrichment of this tracked gas does not depend on
the environment where it resides up to $z\sim2$.

These results further indicate that the bulk of the outskirts ICM was
enriched earlier inside galaxies in the proto-cluster region and then
displaced on larger scales by early AGN feedback.  At $z\gtrsim 2$, in
fact, the potential well of galaxies is
shallower and AGN feedback can be more effective in removing
pre-enriched gas beyond the galaxy boundaries~\cite[see
  also][]{mccarthy2011} and far from active star-formation sites,
contributing to the diffuse enrichment of the inter-galactic medium.
As shown by simulations, AGN feedback has a key role since stellar
feedback alone is not as effective in distributing the enriched
material far from the star formation sites, where it is efficiently
locked back into newly formed stars~\cite[][]{biffi2018}.

\section{Evolution of the ICM metallicity}\label{sec:evol}

The study of the ICM metallicity at different cosmic times is crucial
to confirm the pre-enrichment scenario as well as to to assess the
enrichment history of galaxy clusters.

From the observational side, how much the ICM metallicity evolves with
time is still a matter of debate, although recent X-ray observational
studies suggest a mild evolution in the central regions (especially
for CC systems) and a nearly constant enrichment level at large
cluster-centric distances~\cite[][and
  \Mrev]{ettori2015,mcdonald2016,mantz2017}.
\begin{figure}
  \centering
  \includegraphics[width=0.85\textwidth]{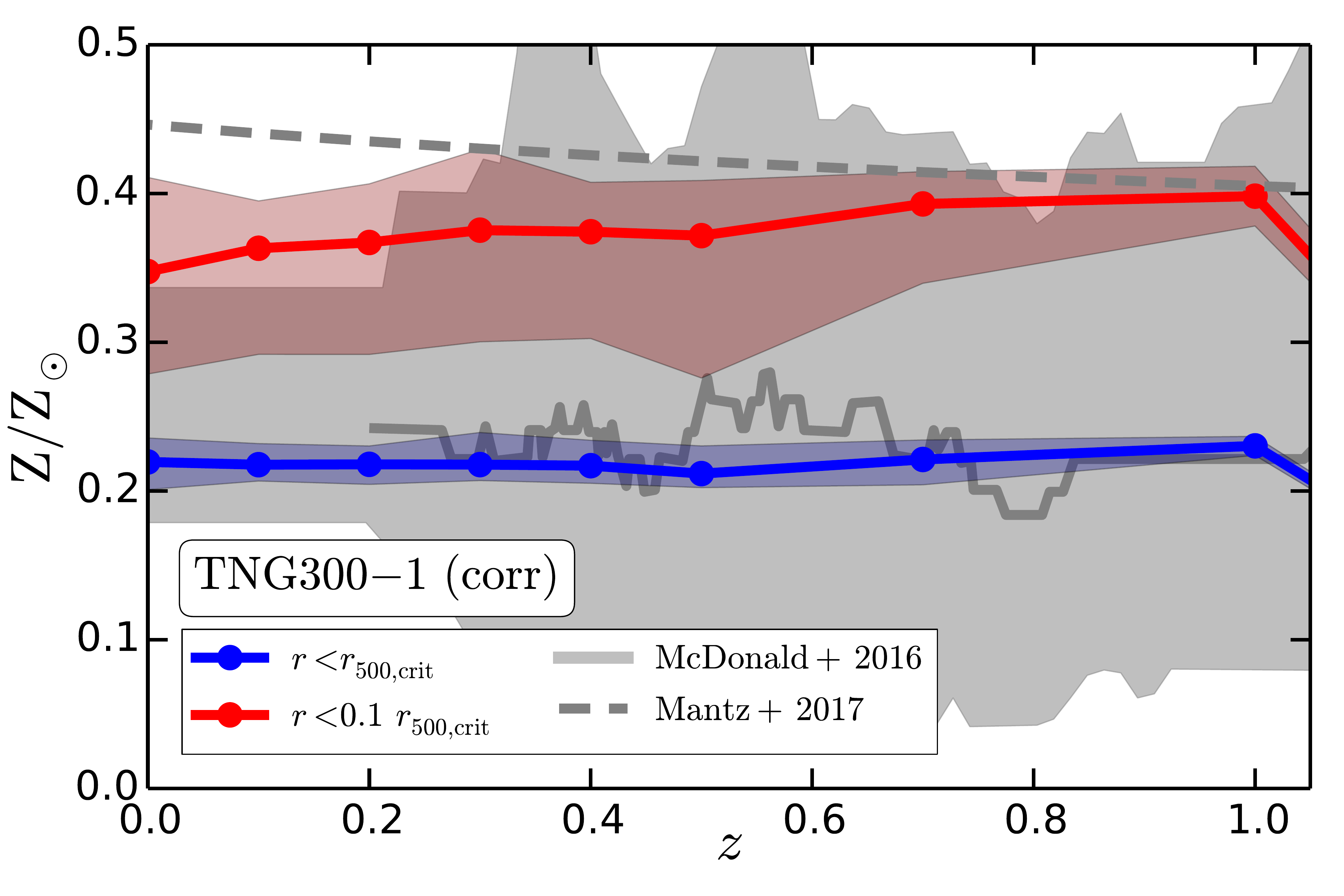}
  \caption{Median redshift evolution of mass-weighted average
    metallicities for clusters in the IllustrisTNG suite; taken
    from~\protect\cite{vogelsberger2018}. Data for two radial ranges
    are reported, inner ($r < 0.1\,\rfive$) and total within $\rfive$,
    and compared against observations by~\protect\cite{mcdonald2016}
    and~\protect\cite{mantz2017}.\label{fig:evol}}
\end{figure}

The ICM pre-enrichment picture, advocated to explain the flat radial
profiles in present-day cluster outskirts found by both X-ray
observations and simulations, is also consistent with the very mild
redshift evolution of the ICM metallicity below $z\sim 2$.

Recent studies based on state-of-the-art simulations accounting a
large variety of physical processes, including AGN feedback, have
shown that the average ICM metallicity and iron-abundance within
$\rfive$ is nearly invariant between $z\sim1.5\mbox{--}2$ and
$z=0$~\cite[][]{biffi2017,vogelsberger2018}. This is consistent with
the flat metallicity profiles at large radii, throughout that redshift
range~\cite[e.g.][]{martizzi2016}. These recent different simulations
all showed that the metallicity profiles present a mild tendency to
become flatter with time, although changes in the normalization are
typically smaller than $\lesssim 10$--$20$\% between $z=2$ and $z=0$.

Concerning the central regions in clusters, an early study based on
simulations without AGN feedback by~\cite{fabjan2008} showed a significant increase of metallicity in cluster central regions ($r<0.2\,R_{180}$) below
redshift $z\sim 1$, with the specific degree of evolution depending on
the assumed IMF. As discussed by those authors, this positive
evolution can actually be the spurious product of the excess of star
formation in the central cluster regions, due to the lack of an
efficient feedback process (only stellar feedback was included in those simulations).
In fact, more recent simulations including also the treatment of AGN
feedback, find an overall weak evolution of the ICM metallicity with
time also in the cluster core.  This is for instance shown
by~\cite{vogelsberger2018}, as reported in Fig.~\ref{fig:evol}, for a
large set of simulated clusters from the recent IllustrisTNG
simulation suite, including both CC and NCC systems. Below $z\sim 1$
simulation data are consistent with no evolution, both in the core
($r< 0.1\,\rfive$) and within the global $\rfive$ region
(respectively, red and blue curves and shaded areas).  From
Fig.~\ref{fig:evol}, we note that the scatter on the median evolution
of the metallicity averages is significantly small for the global
$\rfive$ region: this stresses the universality of the result among
different clusters
and is a further evidence of pre-enrichment, since the global
  mass-weighted value is heavily biased towards the enrichment level
  of the outer regions, where most of the mass is located.
Similar results, were also shown
by~\cite{biffi2017} up to $z\sim 2$. In the cluster core, $r\lesssim
0.1\rfive$, the scatter is instead larger, due to the more important
role played in those regions by astrophysical processes such as
cooling, star formation and feedback, but still it is significantly
smaller than observed~\cite[e.g. compared
  to][]{mcdonald2016,mantz2017}.
A stronger evolution of the core enrichment level, like that reported
by other X-ray works~\cite[][]{ettori2015},
could be ascribed mainly to the evolution of the CC population,
since X-ray-selected samples are biased towards those.

\section{Metal budget in clusters and groups}\label{sec:clus-grp}

Differently from what previously suggested from X-ray observations,
the universality of the ICM enrichment seems to concern not only the
time evolution but also the various cluster scales, ranging from
groups to clusters, as indicated by theoretical and numerical studies.
In the past, X-ray observations of small-mass systems like giant
elliptical galaxies and groups of galaxies typically indicated a lower
enrichment level of the hot gas component with respect to massive
galaxy clusters~\cite[][see also \Mrev, for more details]{fukazawa1998,baumgartner2005,rasmussen2007,rasmussen2009,sun2012,mernier2016,yates2017}, especially in their
cores (10--15\% of $\rfive$). Only recently, these measurements for
the core region have been revised upwards (see results from the CHEERS
sample, \citealt{mernier2018}, and \citealt{truong2018}), confirming
an essentially negligible dependence of the metallicity on the system
temperature.

\begin{figure}
  \centering
  \includegraphics[width=0.7\textwidth]{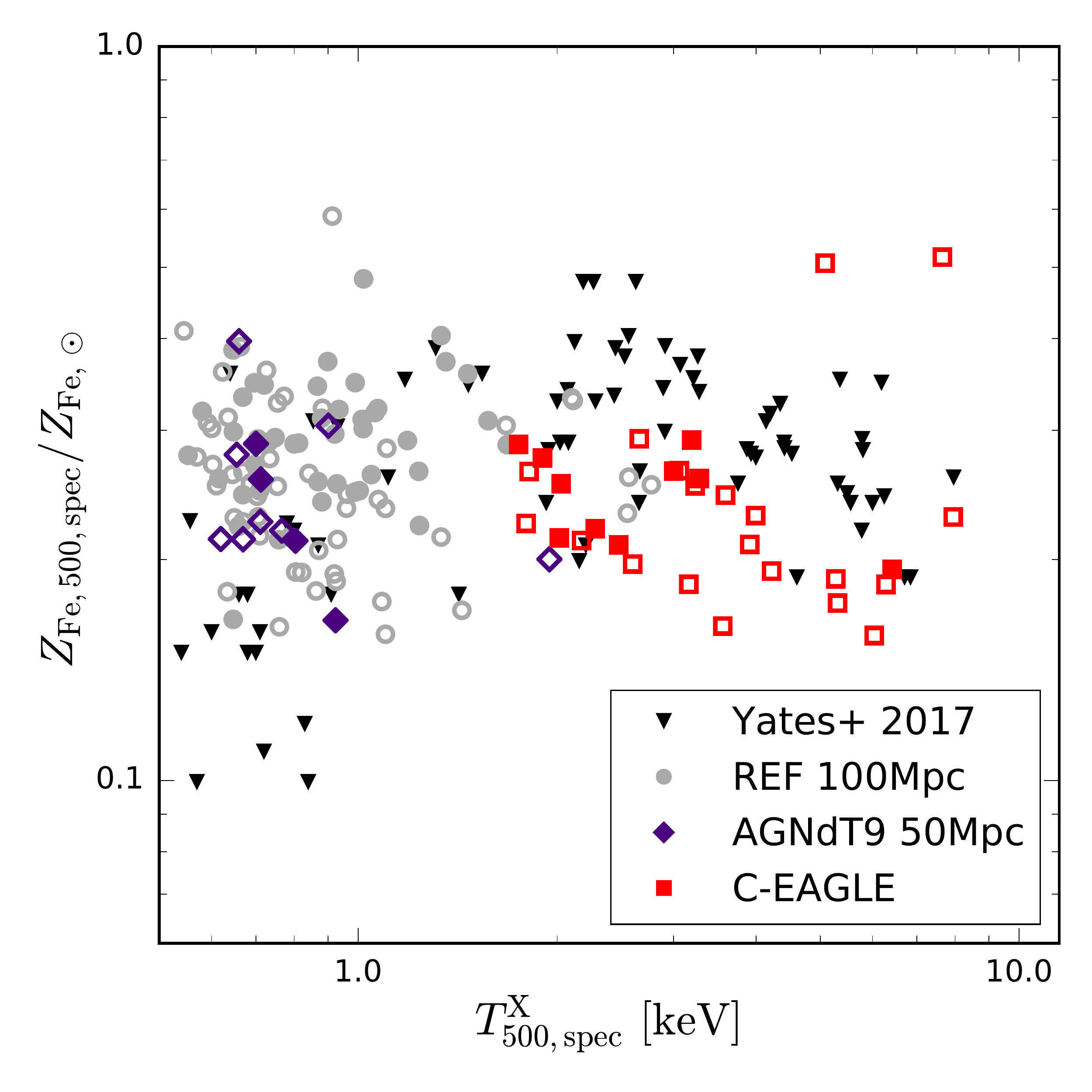}
  \caption{Mass-weighted iron abundance measured within the
    spectroscopic radius $R_{500,spec}$ as a function of the
    spectroscopic temperature, at $z=0.1$, for the C-EAGLE clusters;
    taken from~\protect\cite{barnes2017}. The homogenized observational data
    reported by~\protect\cite{yates2017} are also reported for
    comparison.\label{fig:Fe-T}}
\end{figure}

From the theoretical side, different studies, based both on
simulations~\cite[][]{dave2008,fabjan2010,planelles2014,liang2016,barnes2017,dolag2017,truong2018}
and on semi-analytical models~\cite[][for a recent study]{yates2017},
have shown that the level of enrichment from groups to rich clusters
is essentially similar.
In fact, only a very weak dependence of the metallicity on the system
mass is found, with the intragroup gas slightly more enriched than
the ICM in massive clusters.
This has been shown in a recent study by~\cite{barnes2017}, where the
authors investigate the relation between iron abundance and system
temperature, for the C-EAGLE sample of 30 simulated clusters with
masses in the range $7\times 10^{13} \lesssim M_{500}[M_\odot]
\lesssim 1.2\times 10^{15}$.  In Fig.~\ref{fig:Fe-T} they compare
these results with those from other two sets of EAGLE simulations
including clusters and smaller groups (grey circles and purple
diamonds in the figure), and with consolidated observational data
recompiled by~\cite{yates2017}.  The simulated relation shows good
agreement with observations especially in the range
$T^X_{500,spec}\gtrsim 1\,$keV, whereas the observed iron abundance in
low-temperature groups is found to be lower than in simulations, which
produce a relatively flat mass-weighted metallicity-spectroscopic
temperature relation.

Also \cite{truong2018} found a relatively shallow relation between
metallicity and temperature, from group to cluster scales, both in the
central regions (within 10\% of $\rfive$) and in the global $\rfive$
region. For this study, the authors investigate a large sample of
simulated clusters obtained with an improved, state-of-the-art version
of the SPH code GADGET-3, including the sample of massive clusters for
which the chemical content and enrichment history has been shown to
agree with a variety of observational data
in~\cite{biffi2017,biffi2018}. Simulation results are in good
agreement with recent data by \cite{mantz2017} and from the CHEERS
sample~(\citealt{deplaa2017}, including the updated metallicity
measurements presented by \citealt{mernier2018}).
\cite{truong2018} show that the dependence of gas metallicity on the
system scale is very shallow, basically consistent with being flat,
also when the true mass, instead of temperature, is
considered. Furthermore, they show that the evolution of this relation
with time is negligible between $z\sim 1.5$ and $z=0$.
Despite an average enrichment level similar to clusters, simulations
show that there is a large scatter in metallicities in low-temperature
(low-mass)
systems~(\citealt{dave2008,fabjan2010,planelles2014,liang2016,truong2018}; see
also \citealt{barnes2017}, reported in Fig.~\ref{fig:Fe-T}). This is
related to the diversity of the impact that gas-dynamical processes
and AGN feedback have on the enrichment pattern in smaller systems,
given their shallower potential wells.

\begin{figure}
  \centering
  \includegraphics[width=0.7\textwidth,trim=35 0 10 18,clip]{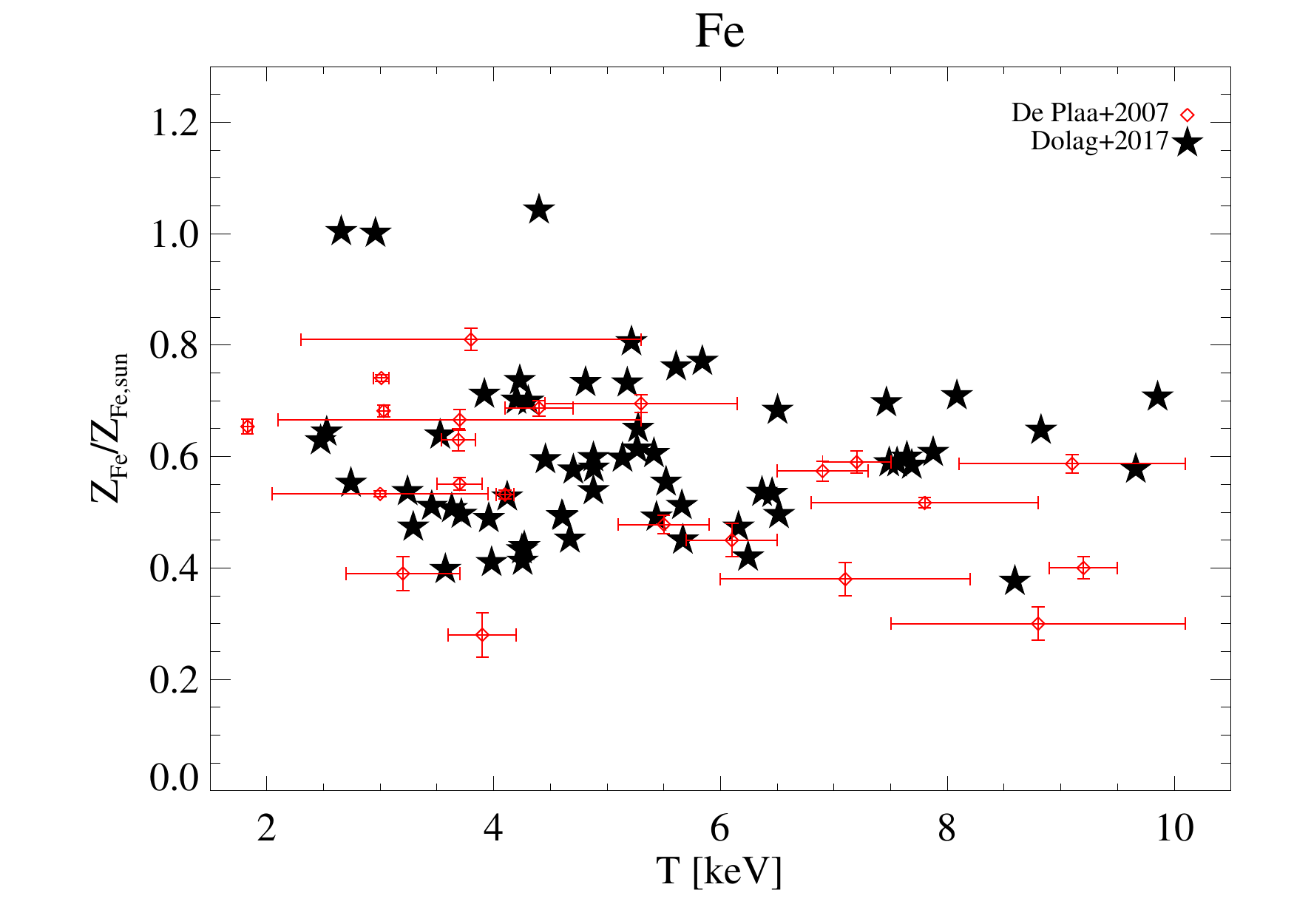}
  \includegraphics[width=0.7\textwidth,trim=15 0 10 15,clip]{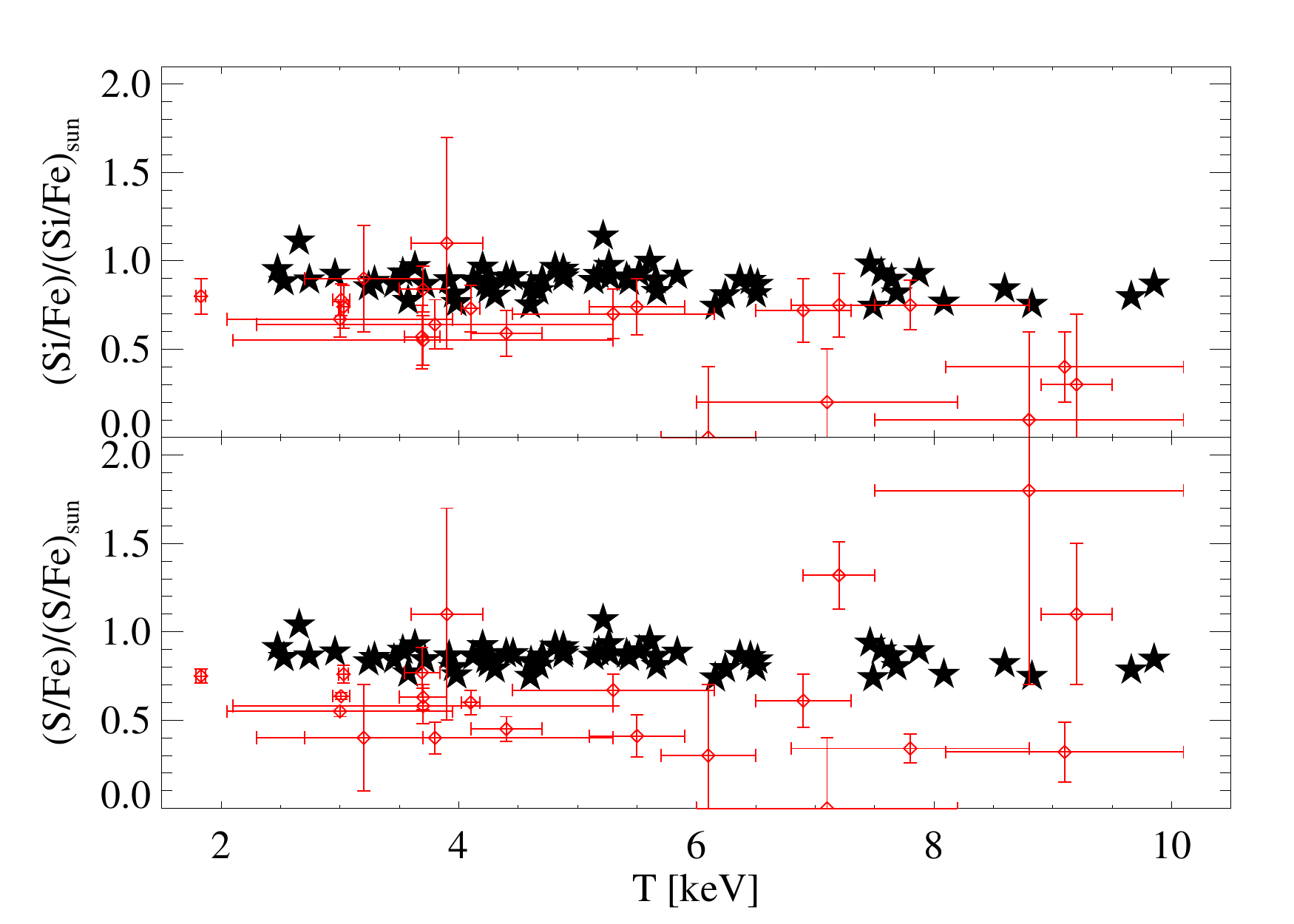}
  \caption{ICM Fe metallicity (top) and abundance ratios (Si/Fe and
    S/Fe; upper and lower insets in the bottom panel, respectively)
    for observed clusters by~\protect\cite{deplaa2007} (red diamonds
    with errror bars) and simulated clusters
    by~\protect\cite{dolag2017} (black stars), as a function of the
    system temperature. All abundances are measured within $R_{2500}$,
    and the temperature is the mean value within
    $R_{500}$.\label{fig:Fe-others-T}}
\end{figure}

In a similar set of simulations from the Magneticum Pathfinder
Project, \cite{dolag2017} investigate the metallicity-temperature
relation in a large sample of simulated massive clusters ($T>2$\,keV),
considering various chemical species separately.  These results are
reported in Fig.~\ref{fig:Fe-others-T}.%
\cite{dolag2017} find essentially flat metallicity-temperature
simulated relations
not only for iron (top panel in Fig.~\ref{fig:Fe-others-T}) but
also for all the other elements considered, with typical values in
agreement with observational data in most of the cases (we show, for
instance, the case of Si/Fe and Si/Fe in the bottom panel of
Fig.~\ref{fig:Fe-others-T}).
Compared to observations by~\cite{deplaa2007}, \cite{dolag2017}
  find differences in the normalization of the simulated  metallicity-temperature
  relation only for a few elements among those followed in the
  simulations, namely for Ne, Ca and Ni, which can nevertheless be
  related to the set of yields adopted.

\section{Current limitations and final remarks}\label{sec:concl}

The study of the ICM chemical enrichment through the cosmic time
provides invaluable information on the evolution of cosmic baryons,
the past star-formation history of clusters and the gas-dynamical
interaction between galaxies and surrounding gas in the overdense
cluster environment.  In this review article, we revised and discussed
the most important results obtained in this field from a theoretical
perspective, with a special focus on numerical simulations.

In particular, the modelling of the chemical properties of the ICM in
state-of-the-art cosmological hydrodynamical simulations has made huge
steps forward in the last years. The description of the main processes
driving the baryonic physics of the hot plasma in groups and clusters
has now reached an unprecedented level of complexity, and simulations
succeed in reproducing a large variety of observational features.  The
emerging global picture indicates that an early enrichment of the gas
in the proto-cluster region must take place in order to explain the
observed (i) uniformity of the enrichment level over large spatial
scales in the outer regions of local clusters and across different
mass scales, from groups to massive systems, and (ii) the time
invariance of the chemical enrichment below redshift $z\sim
1\mbox{--}2$.

Despite their successes, cosmological simulations still need to be
improved in order to consistently reproduce observed chemical and
thermodynamical properties of both clusters and galaxies, accounting
for the various physical processes in place.
The chemical enrichment of cosmic baryons and its evolution are in
fact sensitive to several factors, other than the details of the
chemical evolution modelling itself.
Complex physical and dynamical processes in clusters contribute to mix
and distribute metals over large spatial scales, and to shape the
final ICM enrichment level~\cite[see][for a review]{schindler2008}.
Among these, we recall stellar feedback
in the form of galactic winds, ram-pressure stripping, dynamical
mixing due to large-scale gas motions, buoyancy of gas bubbles
associated to AGN jets, metal diffusion, and depletion of heavy ions
due to dust
formation~\cite[][]{churazov2001,montier2004,rebusco2005,rebusco2006,simionescu2008,simionescu2009,kirkpatrick2015}.
The treatment of these phenomena into the simulations, and their
specific implementation, can therefore impact in a non-trivial way the
chemical properties of the ICM. This was discussed in the
previous sections, in particular, about the various feedback processes accounted for in simulations.

The treatment of dust, for instance, should be certainly included in
future simulations of cosmic structures and galaxy clusters, as it is an important
component involved in the gas cooling and metal depletion.
So far, the effects of dust have been poorly investigated in
cosmological chemo- hydro-dynamical simulations of galaxy
clusters~\cite[see preliminary works
  by][]{aguirre2001,etienne2009,dasilva2009}, being mainly explored
via post-processing computations coupled with numerical simulations of
galaxies and clusters~\cite[e.g.][]{granato2015}.
Recently, a first implementation of the processes affecting the formation
and evolution of dust (in the form of carbonaceous and silicate dust grains)
in SPH simulations has been presented by~\cite{gjergo2018}. In this work,
differences are found with respect to the observed amount of dust in low-redshift
galaxy clusters, although the trend of dust abundances over metallicity in local
galaxies is reasonably reproduced. Further detailed investigations in this direction
are therefore strongly required, for instance to estimate the contribution of dust to the formation of molecular gas,
a primary ingredient for more refined models of star formation.
Moreover, a self-consistent implementation of dust formation and evolution in
cluster simulations is crucial to compute more precisely observational properties
of simulated objects by means of radiative transfer post-processing.
This is essential for multi-wavelength studies of
simulated clusters and of their member galaxies, that can be
compared with a vast range optical and infrared observations. A
successful simulation, in fact, should correctly reproduce both the
large-scale cluster properties and the galactic population as well.

In this respect, a limitation of many current simulations of large
cosmological volumes is still represented by resolution. Increasing the
numerical resolution is a necessary step to consistently resolve galaxies
(i.e.\ the central Brightest Cluster Galaxy as well as satellites) within clusters.
In fact, the study of the processes that determine the chemical
properties of galaxy clusters are strictly linked to the star
formation and metal production histories of cluster galaxies.
The quest for higher resolution is therefore a {\it conditio sine qua
  non} to study the complex interplay between galaxies and surrounding
medium, via energetic and chemical feedback. These processes need to
be captured in great detail especially at high redshift, when galaxies
in the proto-cluster region are likely to enrich and eject gas out to
large distances~\cite[see e.g. the recent study by][]{gupta2018}.

Given the impact of the adopted yields and related parameters (e.g. mass limits for the different stellar phases) on the resulting ICM enrichment, it is crucial that numerical and observational improvements are accompanied by refined calculations of these pillar of stellar evolution models. In fact, the large uncertainties that are still present in these quantities directly affect the normalization of the predicted profiles and global enrichment levels, leaving the accurate comparison between simulations and observations, in terms of absolute values, still unsettled.

\looseness=-1 As demonstrated by the successful results obtained in the recent
years, simulations and observations of the ICM chemical enrichment
must definitely be combined in order to interpret the vast amount of
data that up-coming X-ray telescopes will provide and to improve our
theoretical understanding of the ICM chemo- and thermo-dynamics.
In this respect, we remind that so far ICM chemical abundances of
simulated clusters, both in terms of metal distribution and evolution,
have always been compared against observational data obtained with CCD
cameras on board past and current X-ray telescopes. Nevertheless, the
advent of high-spectral-resolution micro calorimeters on board
next-generation X-ray missions, such as \Athena\ and \XRISM, will
likely open a new scenario with new challenges and further possible
explorations, as it was demonstrated by \Hitomi\ observations of the
Perseus cluster~(\citealp{hitomi2018}; see also the discussion in
\Mrev).

\begin{acknowledgements}
  V.B. is thankful to Elena Rasia and Stefano Borgani for useful
  suggestions that helped improving the manuscript, and to Klaus Dolag
  for kindly providing the simulation data reported in Fig.~10. She
  wishes also to thank Umberto Maio and partial funding support from a
  grant of the German Research Fundation (DFG), number 390015701.
  F.M. is supported by the Lend\"ulet LP2016-11 grant awarded by the
  Hungarian Academy of Sciences. SRON is supported financially by NWO,
  the Netherlands Organization for Scientific Research.
  P.M. acknowledges support from Russian Science Foundation (grant
  14-22-00271).
\end{acknowledgements}

\bibliographystyle{aps-nameyear} 
\bibliography{review}    
\nocite{*}

\end{document}